\documentclass[11pt]{article}
\pdfoutput=1
\usepackage[utf8]{inputenc}
\usepackage{multirow}
\usepackage{amsmath, amsfonts, amssymb}
\usepackage{array}
    \newcolumntype{C}[1]{>{\centering\arraybackslash}m{#1}}
\usepackage{comment}
\usepackage{graphicx}
\usepackage[table]{xcolor}
\usepackage{pdflscape}
\usepackage{psfrag}
\usepackage{amsthm}
\usepackage{enumerate}
\usepackage{arydshln}
\usepackage{pifont}
    
    \newcommand{\emptycell}{$\boldsymbol{\cdot}$}
\usepackage{soul}
\usepackage{slashed}
\usepackage{subcaption}
\usepackage{mathrsfs}
\usepackage{ytableau}
 \usepackage{a4wide}
  \usepackage{tikz}
  \usepackage{tikz-cd}
  \usetikzlibrary{shapes.geometric}
  \usepackage{tcolorbox}
  \definecolor{dark-gray}{gray}{0.20}
  \definecolor{gray}{gray}{0.30}
  \definecolor{light-gray}{gray}{0.80}
  \definecolor{dark-red}{rgb}{0.7,0,0}
  \definecolor{dark-green}{rgb}{0.1,0.4,0}
  \definecolor{dark-blue}{rgb}{0.3,0.3,0.7}
  \definecolor{light-blue}{rgb}{0.8,0.8,1}
      \definecolor{swamp}{RGB}{240, 199, 197}

\newcommand{\xmark}{\ding{55}}

\usepackage{newunicodechar} 
\usepackage{setspace}

\usepackage{ifthen}
\renewcommand*{\arraystretch}{0.9}
\usepackage{longtable}

\newcommand{\be}{\begin{equation}}
\newcommand{\ee}{\end{equation}}
\newcommand{\eq}[1]{(\ref{#1})}

\def\be{\begin{equation}}
\def\ee{\end{equation}}
\def\bea{\begin{eqnarray}}
\def\eea{\end{eqnarray}}

\newcommand{\R}{\mathbb{R}}

\newcommand{\Z}{\mathbb{Z}}
\AtBeginDocument{

}

\newcommand{\ed}{{\rm d}}

\newcommand{\figref}[1]{figure~\ref{#1}}

\numberwithin{equation}{section}

\usepackage{jheppub}

\usepackage{cleveref}

\hypersetup{
	colorlinks=true,
	linkcolor=dark-blue,
	citecolor=dark-red,
	urlcolor=dark-green,
	linktoc=page
}

\theoremstyle{definition}

\theoremstyle{remark}

\crefname{appendix}{Appendix}{Appendices}

 \title{dS no-go's for Riemann-flat compactifications and semidefinite optimisation}
\title{de Sitter no-go theorems and semidefinite optimisation}
\title{dS minima on Riemann-flat compactifications: No-gos and semidefinite optimisation}
 \title{de Sitter no-go's for Riemann-flat manifolds\\ and a link to semidefinite optimisation}

\author{Bruno Valeixo Bento} 
\author{and Miguel Montero}
\affiliation{Instituto de F\'{i}sica Te\'{o}rica IFT-UAM/CSIC,
C/ Nicol\'{a}s Cabrera 13-15, Campus de Cantoblanco, 28049 Madrid, Spain}

\emailAdd{bruno.bento@ift.csic.es}
\emailAdd{miguel.montero@csic.es}

\abstract{We establish a no-go theorem in the context of string and M-theory flux compactifications on Riemann-Flat manifolds with Casimir energy. Specifically, we show that no dS minimum exists in this setup in dimension $d>3$. The case of dS$_3$ minima is not excluded, but their actual fate can only be ascertained via an explicit construction. We also point out that the problem of finding dS minima on RFM's and more general flux compactifications is mathematically equivalent to a semidefinite programming problem, identical to those studied in  CFT bootstrap, and hence the search for dS can benefit from the existing vast literature and numerical tools. We illustrate this in a toy model.}

\begin{document}
\emergencystretch 3em
\hypersetup{pageanchor=false}
\makeatletter
\let\old@fpheader\@fpheader
\preprint{IFT-25-115}

\makeatother

\maketitle

\hypersetup{
    pdftitle={},
    pdfauthor={},
    pdfsubject={}
}

% Notation commands
\newcommand{\D}[1][\gamma]{\mathbf{D}_{\bf #1}}
\newcommand{\bvec}[1][\gamma]{\vec{b}_{\bf #1}}
\newcommand{\RFM}[2][T]{%
  \ifthenelse{\equal{#1}{T}}%
    {\frac{T^{#2}}{\Gamma}}%
    {\frac{\mathbb{R}^{#2}}{\mathcal{B}}}%
}
\newcommand{\Tr}[2]{{\rm Tr_{\bf #1}}(#2)}
\newcommand{\TrB}[1]{\Tr{B}{#1}}
\newcommand{\TrF}[1]{\Tr{F}{#1}}
\newcommand{\Vcas}{V_{\text{Cas}}}

\section{Introduction and Conclusions}
\label{sec:introduction}

Whether String Theory can harbor long-lived accelerated expansion remains to this day an open question. Answering this question is not only crucial in order to connect to observations \cite{SupernovaCosmologyProject:1998vns,SupernovaSearchTeam:1998fmf}, but also essential in the context of the non-supersymmetric Landscape. In fact, it has long been argued that accelerating cosmologies are not expected near the boundaries of moduli space, where strings are arbitrarily weakly coupled or we have some other control parameter that can be tuned arbitrarily \cite{Dine:1985he,Ooguri:2006in,Lee:2019wij}. Paired with the lack of explicit string constructions, these have led to concrete Swampland conjectures \cite{Palti:2019pca,vanBeest:2021lhn} that acceleration may not exist in asymptotic regions of moduli space \cite{Danielsson:2018ztv,Obied:2018sgi,Ooguri:2018wrx,Andriot:2018wzk,Andriot:2018mav,Bedroya:2019snp}, for which there is now ample systematic evidence (see e.g. \cite{Grimm:2019ixq,Calderon-Infante:2022nxb,Etheredge:2024tok}).

As a consequence, any stringy realisation of long-lived phases of accelerated expansion is pushed away from the asymptotic boundaries of moduli space. In the absence of a fully non-perturbative gravitational setup with any measure of control, accelerated expansion scenarios usually end up living on the edge between asymptotic and non-perturbative regimes, where a perturbative description in terms of weakly-coupled strings or a large internal space may still hold, but where additional non-asymptotic ingredients come into play, avoiding the asymptotic no-go theorems. Existing de Sitter scenarios, such as KKLT \cite{Kachru:2003aw} and LVS \cite{Balasubramanian:2005zx}, fall in these category and remain under debate (see e.g. \cite{Gao:2020xqh, Demirtas:2021nlu,Junghans:2022exo,Bena:2022cwb,Gao:2022fdi,Lust:2022lfc,Lust:2022xoq,Hebecker:2022zme,Bena:2022ive,ValeixoBento:2023nbv,McAllister:2024lnt,Kim:2024dnw,Moritz:2025bsi}) largely due to the difficulty to control the mix of ingredients that are required. This complexity can be understood in part through one particular no-go theorem \cite{Gibbons:1984kp,Maldacena:2000mw}, which excludes de Sitter solutions with only classical supergravity ingredients (fluxes and curvature). In an attempt to escape this, existing dS scenarios typically include stringy sources such as orientifolds, non-perturbative effects, and/or higher-derivative corrections, which are beyond the reach of 10 or 11-dimensional supergravity. This makes them intrinsically complicated and leads to questions e.g. regarding the backreaction of these stringy singular sources \cite{Gao:2020xqh}. 

In \cite{ValeixoBento:2025yhz} we have initiated the study of flux compactifications on non-supersymmetric Riemann-flat manifolds (RFM's) with Casimir energy (along the lines pioneered in \cite{DeLuca:2021pej}) and constructed an explicit $dS_5$ maximum solution in M-theory. The advantage of this setup is that all ingredients can be properly described within supergravity, saving us from the complications inherent to the stringy ingredients of other settings. Not only is this construction simpler as a whole, it also lends itself to easier scrutiny that could quickly determine the fate of any particular solution. In this paper, we determine under which conditions a $dS$ \emph{minimum} (i.e. metastable dS vacuum) supported only by flux and Casimir energies on Riemann-flat manifold compactifications can exist. As we will see, this setup is extremely constrained and we will be able to narrow down the possibilities quite significantly. 

We are interested in a compactification of a $D$-dimensional supersymmetric theory on a $k$-dimensional RFM down to $d$ dimensions. From the lower-dimensional EFT perspective, a de Sitter solution corresponds to a positive minimum of the scalar potential for all the scalar fields in the theory. The set of scalar fields in the EFT includes all scalars obtained through dimensional reduction of the higher-dimensional theory---moduli of the higher-dimensional theory itself, geometric moduli related to the RFM and scalars that descend from form fields with components along the internal space. Among these fields is typically the overall volume $\mathcal{V}\sim R^k$ of the compact space that becomes a modulus of the low-energy theory\footnote{There are (non-geometric) backgrounds with no volume modulus, such as Type IIB backgrounds with no K\"ahler moduli that are mirror dual  of Type IIA on rigid Calabi-Yau manifolds \cite{Vafa:1991uz,Candelas:1993nd}. These can be described by Landau-Ginzburg models \cite{Vafa:1989xc}, and have been used both in the search for dS solutions and for moduli stabilisation more generally \cite{Becker:2006ks,Bardzell:2022jfh,Becker:2022hse,Cremonini:2023suw,Becker:2024ijy,Chen:2025rkb}.}.

A metastable vacuum requires all these scalars to be \emph{stabilised}, i.e. it requires a \emph{minimum} along every direction in field space. This means that in order to show that a dS vacuum exists, one must show that a minimum is present for all moduli. However, excluding a dS minimum is easier---one can do it by showing that a minimum \emph{does not} exist along a single specific direction. Since the volume modulus is always present in geometric settings, if one cannot find a minimum for the volume, a dS minimum solution is automatically excluded. 

A scalar potential for the moduli can arise from various sources. At the classical level, a potential is generated by non-trivial fluxes threading the compact space or by its curvature. Perturbative and non-perturbative corrections to the leading order theory can also contribute to the potential of certain moduli, as well as localised sources such as D-branes or orientifold planes, both of which requiring string theory for a proper description. 

In order to decide whether these contributions can lead to a minimum along the volume direction, one must check their volume dependence and their sign. Each contribution to the potential will behave as a power of $R$ and, writing the $d$-dimensional EFT in Einstein frame, the power will necessarily be negative---this implies that all contributions to the potential vanish as $R\to\infty$ consistently with the fact that they arise through compactification. Therefore, a minimum along the $R$ direction requires contributions with different signs.

\begin{figure}
    \centering
    \includegraphics[width=0.75\linewidth]{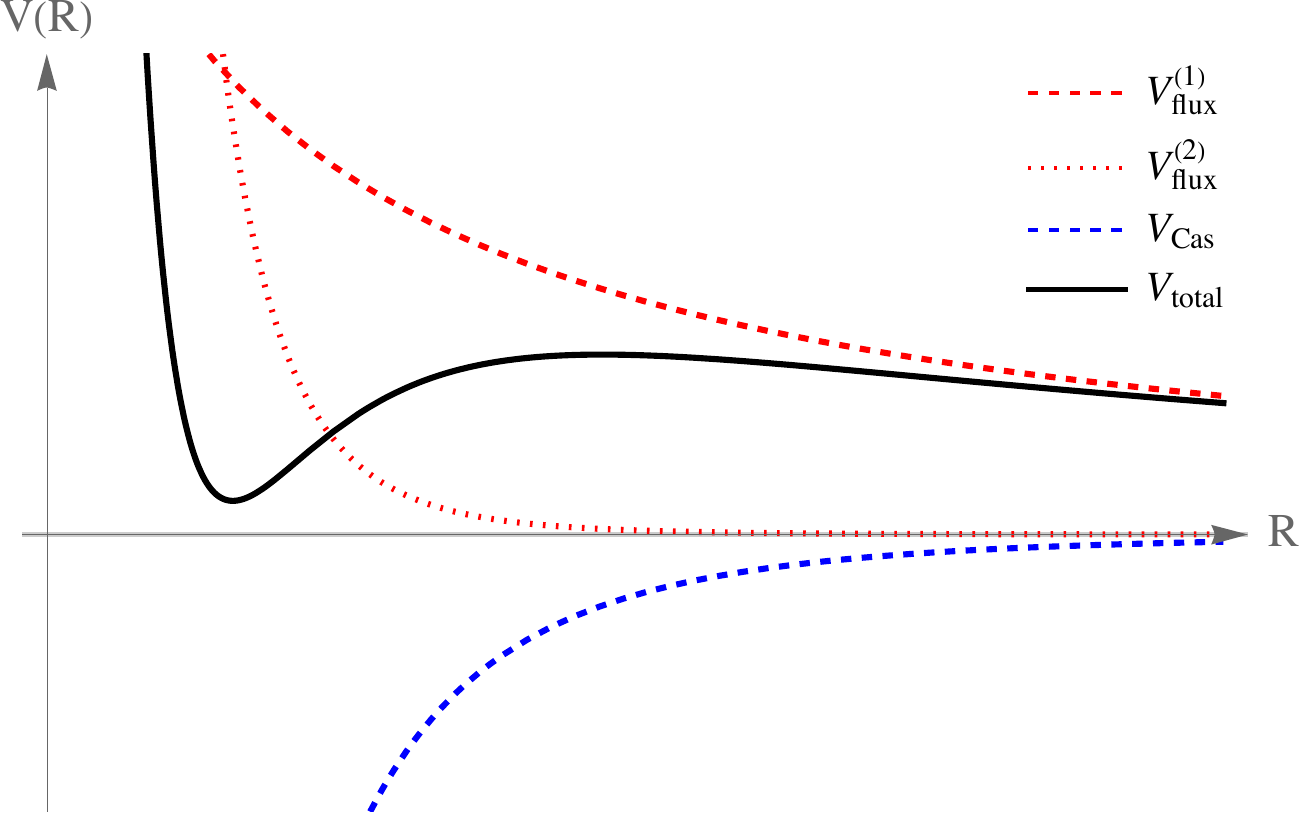}
    \caption{ In this paper, we look for dS minimum solutions where a negative Casimir energy contribution (blue dashed line) and, at least, two flux contributions (red lines) generate a combined potential (black solid line) for the volume modulus $R$ with a local positive metastable minimum. There is no curvature contribution since the compactification manifold is Riemann-flat. Due to Einstein frame rescalings, the value of the minimum can easily become small in Planck units even for moderately large $R$.}
    \label{fig:key-idea}
\end{figure}

This simple reasoning leads us to the typical structure of a potential with a dS minimum: a potential with (at least) three terms, one of which contributes with a negative sign (see Fig.\ref{fig:key-idea}). The negative term can arise in different ways: it could come from positive curvature of the internal space; from localised O-plane sources; or e.g. from a negative Casimir energy. In this paper, we want to focus our attention on the latter, by compactifying on Riemann-flat manifolds and avoiding localised stringy sources like O-planes. Our negative contribution to the potential will thus arise from Casimir energies sourced by the $D$-dimensional massless free fields on the Riemann-flat background. In fact, the Casimir energy has a universal volume dependence and it will necessarily scale as $\Vcas\sim R^{-d}$ in the EFT scalar potential, therefore contributing with a single negative term as far as the volume modulus $R$ is concerned. In \cite{ValeixoBento:2025yhz} we show how to compute these Casimir energies explicitly on RFMs, not only to obtain $\Vcas$ at the level of the EFT, but also the Casimir energy density profile on the RFM, which can be used to solve the $D$-dimensional equations and check whether any solution of the EFT is indeed a good solution of the $D$-dimensional theory.  

In the minimal setup that includes flux and Casimir contributions only, the scalar potential for $R$ takes the form
\begin{equation}
    \frac{V^{(d)}}{M_d^{d}} \propto \frac{1}{R^{\frac{d}{d-2}k}}\left\{\sum_{p}\frac{n_p^2}{R^{2p-k}} - \frac{\mathcal{C}}{R^d}\right\} \,,
    \label{eq:potential-R}
\end{equation}
where each term in the sum over $p$ arises from a $p$-form flux, and we defined the Casimir coefficient $\mathcal{C}$ in such a way that makes the negative sign explicit. A favourable Casimir contribution must have $\mathcal{C}>0$, which is interpreted as a surplus of boson over fermion degrees of freedom in the low-energy theory. 
Noting that a dS minimum requires the Casimir term to appear between two flux terms with respect to their volume scaling, such that a flux on the left brings the potential back up as $R\to 0$ and a flux on the right brings it to zero from above as $R\to\infty$ (see Fig.\ref{fig:key-idea}), it immediately follows that at least two flux terms $p_1,p_2$ satisfying
\begin{equation}
    2p_1 < D=d+k < 2p_2 
    \label{eq:constraint-D}
\end{equation}
are required for a dS minimum to exist, with the inequalities being strict. Equation \eqref{eq:constraint-D} is extremely constraining, particularly since $p_1,p_2 \leq k$ for internal fluxes threading the $k$-dimensional compact space. Indeed, this implies 
\begin{equation}
    D < 2p_2 \leq 2k = 2(D-d) \implies D>2d \,,
    \label{eq:constraint-Dd}
\end{equation}
which becomes a very strong constraint on our setup. It means that there can be no minimum for $d>5$; in fact, we can go further and also exclude the $d=5$ case, since a minimum would require a supersymmetric theory in $D=11$ with a 6-form flux, but M-theory (the only supergravity theory in $D=11$) has no such flux. In the following sections, we will follow this procedure of looking for appropriate fluxes in $D$-dimensional supersymmetric theories and determine which ones can be excluded due to the absence of these fluxes. For now, we can already state that 
\begin{center}
    \emph{no Casimir-flux dS minimum exists for $d\geq 5$}
\end{center}
for Riemann-flat compactifications of supersymmetric theories. Our analysis here is very similar to the one performed in \cite{Parameswaran:2024mrc}, where other no-go theorems are found like this. 

Although $dS_2$ may also be quite interesting for connection to e.g. JT gravity and matrix models \cite{Mertens:2022irh,Maldacena:2019cbz,Iliesiu:2020zld,Nanda:2023wne}, here we will focus on $d=3,4$. With the requirement that the $D$-dimensional theory be supersymmetric in order to regularise Casimir energies, and demanding that it admits a Minkowski vacuum, we are restricted to 
\begin{align*}
    &D = 9,10,11 \quad \text{for}\quad d=4 \,, \\
    &D = 7,8,9,10,11 \quad \text{for}\quad d=3 \,.
\end{align*}
In all these cases the minimal amount of supersymmetry is 16 supercharges. 
A systematic analysis of all $D$-dimensional supersymmetric theories satisfying these constraints will allow us to determine which ones---if any---might harbor de Sitter minima in this setup. The goal of this paper is to perform such analysis, which we do in the following sections. 

We will find the following no-go theorems:
\begin{itemize}
    \item[\textbf{No-go:}] \emph{There is no dS$_d$ minimum in Casimir-flux compactifications on Riemann-flat manifolds for $d>4$}.
    \item[\textbf{No-go:}] \emph{dS$_4$ minima is ruled out for all cases starting with a Minkowski theory in either 16 or 32 supercharges}.
\end{itemize}
In contrast, dS$_3$ minima are possible, in principle, for supergravity theories with both 16 and 32 supercharges. 

It is worth emphasising that these results apply to dS minima (i.e. metastable de Sitter vacua); by relaxing the requirement that the solution be a minimum of the potential, allowing for example for dS maxima, the setup is much less constrained (cf. dS$_5$ maximum from M-theory \cite{ValeixoBento:2025yhz}). Hence many of the cases that are ruled-out for dS minima by our analysis may still allow for dS maxima, or even quintessence-like models, that may be phenomenologically interesting. Nonetheless the results in this paper illustrate one of the (many) advantages of RFM compactifications with Casimir energies: not only is the setup itself extremely constrained, but together with the possibility of fully detailed analyses of specific examples \cite{ValeixoBento:2025yhz}, it allows us to explore this corner of the Landscape in an explicit and conclusive way. Moreover, the no-go results we find also narrow down our search for concrete de Sitter vacua, which we will continue in future work.

This note is structured as follows: Section \ref{s2} contains the proofs of the no-go theorems quoted above, together with their assumptions. Section \ref{sec:comments-dS3} explains how to recast the search of dS minima with Casimir energies on RFM's as a semidefinite bootstrap program. Section \ref{s4} contains some brief remarks about the results and our thoughts on what to do next.

\section{No-gos for SUGRA}\label{s2}

Motivated by the strong constraints on Casimir-flux de Sitter vacua discussed in the previous section, we will now perform a systematic analysis of supergravity theories in dimensions $D=7,8,9,10,11$, with 16 and 32 supercharges for which both field content and kinetic terms are fixed by the amount of supersymmetry \cite{Weinberg:2000cr,Freedman:2012zz}. This will allow us to rule out many of them either due (i) to the lack of appropriate $p$-form fluxes to turn on or (ii) to higher-dimensional moduli (in particular, the dilaton modulus) that cannot be fixed with only the ingredients considered here. 
We summarise the results in Tables \ref{tab:theories-32supercharges} and \ref{tab:theories-16supercharges}.

Let us begin! Consider the action of a $D$-dimensional theory with a higher-dimensional modulus $\phi$ written in Einstein frame \cite{Heidenreich:2015nta},
\begin{align}
    2\kappa_D^2\,S^{(D)} \supset \int d^Dx\,\sqrt{-G}\,\left[\left(\mathcal{R}_D - \frac12(\nabla\phi)^2\right)
    -\frac{e^{-\alpha_p\phi}}{2}F_{p}^2
    + \rho_{\rm Cas}\right] \,,
\end{align}
where we include the $\phi$-independent contribution from Casimir energies and make explicit the coupling between $\phi$ and the field-strength $F_p$. Due to this coupling, the appropriate dual field-strength is $F_{D-p}\propto e^{-\alpha_p\,\phi}\star F_p$ \cite{Heidenreich:2015nta}. Upon dimensional reduction, these forms contribute to the scalar potential of the $d$-dimensional action through non-trivial fluxes that must be appropriately quantised \cite{Polchinskiv2}, which for $\phi$ takes the form\footnote{In a slight abuse of notation we still call $\phi$ the $d$-dimensional scalar obtained from dimensional reduction of the $D$-dimensional scalar $\phi$.}
\begin{equation}
    V^{(d)}(\phi)\propto \sum_p \frac{f_p^2}{e^{\alpha_p\phi}R^{2p}} - \frac{\mathcal{C}}{R^{D}} \,.
    \label{eq:dilaton-potential}
\end{equation}
We have not included an overall integration over the compact space, which produces a volume factor of $R^k$, nor the Einstein frame rescaling in lower dimensions; these will be immaterial for the considerations here and would only clutter the equations.
Although we write the sum over all flux terms (electric and magnetic duals) to parallel the potential \eqref{eq:potential-R}, we should keep in mind that magnetic dual fluxes have a $\phi$-dependence related to their electric counter-parts. The relationship is $\alpha_{D-p}=-\alpha_p$, due to electric-magnetic duality. Just like for $R$, the flux contributions to the $\phi$ potential are always positive, which means that a dS minimum along the $\phi$ direction requires the same minimal 3-term structure with the negative Casimir term in-between two flux contributions. This translates into the requirement that there are at least two fluxes $p_1,p_2$ satisfying 
\begin{equation}
    \alpha_{p_1} < 0 < \alpha_{p_2} \,.
    \label{eq:constraint-dilaton-powers}
\end{equation}
In particular, \eqref{eq:constraint-dilaton-powers} means that we need both positive and negative powers of $e^\phi$ in order to stabilise it in a dS minimum with fluxes. Any setup in which we are not allowed to turn them on cannot rely on fluxes to stabilise the dilaton. 

In the context of flux compactifications on RFM's, there is another way to fix these moduli, which we shall call \emph{duality freeze-out}. In some cases the $D$-dimensional theory will contain some duality group under which the scalars are charged. whenever this happens, we are allowed to turn on non-trivial bundles for the duality group that can fix these moduli at a given value, in a discrete version of the Scherk-Schwarz mechanism \cite{Scherk:1978ta}. In more detail, when compactifying the theory on the $k$-dimensional RFM, the discrete bundle corresponds to turning on a holonomy around  one-cycles in the compactification manifold $\mathcal{M}_k$, which introduces boundary conditions that effectively stabilise the moduli at the KK scale. 

However, we will not always be able to use duality freeze-out, even when a suitable duality group is present. Since some $p$-forms are also charged under these dualities, turning on a non-trivial bundle for them can end up projecting out the fluxes that we need for the $R$ stabilisation in the first place (see detailed discussion for Type IIB below). Ordinary fluxes are classified by ordinary cohomology of the underlying manifold; thus, a $p$-form flux can only be switched on if $H^p(\mathcal{M}_k,\mathbb{Z})$ is nontrivial. For fields charged under a discrete duality symmetry group $\Gamma$, the appropriate object is cohomology with $\Gamma$-twisted coefficients  $H^p(\mathcal{M}_k,\Gamma)$. These can be computed via standard techniques (see \cite{Aharony:2016kai,Etheredge:2023ler} for some examples in the context of string theory), but for RFM's it is often easiest to determine them via the covering $T^k$; specifically,  the free part of $H^p(\mathcal{M}_k,\Gamma)$ is represented by $p$-forms on $T^k$ that are invariant under the combined action of $\Gamma$ and the isometry group that generates the RFM.

With these preliminaries, we are ready to study the theories of interest. As explained above, these all have $D\geq 7$, and we focus on supersymmetric theories with a Minkowski vacuum so that the techniques of \cite{ValeixoBento:2025yhz} for computing Casimir energies reliably on RFM's can be used. This means that we only discuss theories with 16 or 32 supercharges. Most of these theories have moduli, whose stabilization must be accounted for when finding a dS minimum. In many (but not all) of the models, there can be one or more scalars that are not charged under the duality group and hence cannot be dealt with by duality freeze-out. A prototypical example is the dilaton in $D=10$ type IIA supergravity.  In some other theories, like type IIB in ten dimensions, all scalars can be frozen via dualities. In cases like this, we will assume that these scalars pose no obstacle in the construction of dS minima, and can be safely ignored. Actually, a lot more work would be needed in a particular example to see that this is indeed the case (e.g. the RFM must support the right bundle). So we are considering the best case scenario for each theory; as we will see, this will not be enough. We have arranged the discussion by whether there are scalars uncharged under duality symmetries or not; we will now discuss each case in turn.

\subsection{Theories with uncharged scalars under the duality group}
\label{sec:uncharged}
All theories with 16 supercharges, and some with 32, fall into this class. We will study both cases.

\subsubsection{16 supercharges}
\label{sec:uncarged-16}
Let us first consider theories with 16 supercharges. These exist only for $D\leq 10$, and only admit gravity and vector multiplets \cite{Freedman:2012zz}. The gravity multiplet contains the metric, $(10-D)$ vector fields, a 2-form $B_{2}$ and a scalar $\phi$. Although these theories admit rich duality groups, as we review below, $\phi$ is always uncharged, so it must be stabilised by fluxes. As for the fluxes that one can turn on, there is a $3$-form flux coming from the field-strength of $B_2$, and $2$-form fluxes from vectors, as well as the magnetic dual $(D-3)$ and $(D-2)$-form fluxes\footnote{In  $D=10$ the fluxes associated to the vectors are absent since there are none.}. 

Let us first address what ingredients would be required to stabilize the volume modulus $R$ of the RFM we use as compactification manifold. Condition \eqref{eq:constraint-Dd} implies that a dS$_4$ minimum requires a $p_2$ flux with $p_2 > D/2 > 4$, which means this role must be played by one of the magnetic dual fluxes. The further requirement that the fluxes be internal, $p_2 \leq k = D-d$, when applied to the available $(D-2)$ and $(D-3)$-form fluxes, implies $d\leq 3$, with $d=3$ requiring $H_{D-3}$ flux from the gravity multiplet 2-form. Thus, from considering the $R$ stabilisation alone, we conclude that theories with 16 supercharges do not have Casimir-flux dS$_4$ minimum solutions.
If a Casimir-flux dS$_4$ minimum exists, it must come from a theory with 32 supercharges.

\begin{itemize}
    \item[\textbf{No-go:}] \emph{Theories with 16 supercharges have no Casimir-flux dS$_4$ minimum solutions}. 
\end{itemize}

Nevertheless, theories with 16 supercharges can in principle harbor dS$_3$ minima, which are necessarily supported by $H_{D-3}$ flux. With this, the volume modulus can in principle be stabilised. 
However, in addition to the scalar $\phi$ in the gravity multiplet which we just discussed, theories with sixteen supercharges also have  $(10-D)$ scalars in each vector multiplet. The number of vector multiplets itself is different in different components of the moduli space, with the largest number being $26-D$. A de Sitter minimum requires that these are stabilised too. Fluxes can stabilize them in principle, but now we can also use duality freezing---the moduli in the vector multiplets live in the Narain coset \cite{Narain:1985jj,Narain:1986am}
\begin{equation}
    \frac{SO(10-D,n_V,\R)}{SO(10-D,\R)\times SO(n_V,\R)}/\Gamma \,,
    \label{eq:Narain-coset}
\end{equation}
where $\Gamma$ is the automorphism group of the charge lattice for the vectors in the $n_V$ vector multiplets \cite{Narain:1985jj,Narain:1986am}. In order to freeze the moduli through non-trivial bundles we must look for fixed points that correspond to lattice isometries \cite{Giveon:1988tt,Giveon:1994fu,Fraiman:2018ebo}, i.e. that can be rewritten as a lattice change of basis. These correspond to crystallographic groups of the lattice. Duality freeze-out for these scalars will be possible if we find symmetries that can fix all moduli. This is definitely possible at least in some components of moduli space.  For instance,  for the so-called AOB background \cite{Aharony:2007du}, which is a 9-dimensional theory with just a single vector multiplet, the charge lattice is $\Lambda^{AOB}=\Gamma_{1,1}$, the unique even, self-dual lattice in signature $(1,1)$. Its automorphism group is $\mathbb{Z}_2$, which is a T-duality transformation and fixes the single scalar in the vector multiplet. For $D\leq 8$, the duality group of all known theories contains nontrivial elements that allow one to fix several scalars. In particular, for $D=7$, there are again theories with just one modulus, which can therefore be fixed; for $D=10$ there are no moduli. Thus, at least in the cases $D=7,9,10$, de Sitter vacua can potentially exist in three dimensions with all vector multiplet moduli stabilised by duality freeze-out, and therefore dS$_3$ minima are not excluded in this scenario\footnote{Again we emphasize that not ruling out dS does not mean it exists in a given setup. For instance, a dS$_3$ from $D=7$ would require a  four-dimensional RFM, of which there are only 84 \cite{lambert2025closed}. The landscape here is so small that it could very well be that no viable possibility exists.}.

\subsubsection{32 supercharges}
\label{sec:uncharged-32}
 Some theories with 32 supercharges also have a dilaton-like modulus and no duality group that can freeze it. Theories with 32 supercharges only have a single multiplet, the gravity multiplet, and at low energies---in the supergravity limit---the action and matter content are uniquely fixed. The two cases with uncharged scalars are $D=9,10$. We will discuss each in turn.

\subsection*{\ding{71} $D=10$}

The action of Type IIA supergravity is
\begin{align}
    2\kappa_{10}^2\,S_{\rm IIA} =\,& \int d^{10}x \left(R - \frac12(\partial\phi)^2 - \frac{e^{-\phi}}{2}|H_3|^2 - \frac{e^{\frac32\phi}}{2}|F_2|^2 - \frac{e^{\frac12\phi}}{2}|F_4|^2 \right) \nonumber \\
    & -\frac12\int B_2\wedge F_4\wedge F_4 \,.
    \label{eq:IIA-action}
\end{align}
It includes a scalar, the dilaton $\phi$, which must be stabilised. There is however no duality group under which $\phi$ is charged, so we must do it via fluxes. The structure is now more complicated than in 16 supercharges, since there are many more fluxes available as is clear from the action.

As before, let us first check the volume modulus. 
Type IIA has forms allowing for fluxes with $p=2,3,4,6,7,8$, including magnetic duals. According to \eqref{eq:constraint-Dd}, a dS$_4$ minimum requires a $p_2=6$ flux  that does exist in Type IIA, as well as a $p_1<5$ flux whose role can be played by either $F_2$, $H_3$ or $F_4$. Therefore, we cannot immediately exclude a dS$_4$ minimum in Type IIA based on \eqref{eq:constraint-Dd} alone. 

We still have the dilaton to stabilise. 
At first sight we seem to have sufficient dilaton-flux terms to satisfy constraint \eqref{eq:constraint-dilaton-powers}, so that again a dS$_4$ minimum could not be immediately ruled out. 
Yet a more convenient definition of the moduli does allow one to rule out Type IIA. Take the potential for $\phi$ and $R$,
\begin{equation}
    V^{4d}(\phi,R) \propto \frac{e^{-\phi}}{2}\frac{n_3^2}{R^{12}} + \frac{e^{\frac{5-p}{2}\phi}}{2}\sum_p \frac{n_p^2}{R^{2p+6}} - \frac{\mathcal{C}}{R^{16}} \,,
\end{equation}
and consider the field redefinition $R\to e^{-\frac{\phi}{4}}\tilde{R}$. In terms of this rescaled volume modulus, the potential becomes 
\begin{equation}
    V^{4d}(\phi,\tilde{R}) \propto e^{4\phi}\left(\frac{e^{-2\phi}}{2}\frac{n_3^2}{\tilde{R}^{12}} + \frac{1}{2}\sum_p \frac{n_p^2}{\tilde{R}^{2p+6}} - \frac{\mathcal{C}}{\tilde{R}^{16}}\right) \,,
\end{equation} 
which makes it manifest that we do not have a three-term structure for the $\phi$ scalar potential. A dS$_4$ minimum is thus ruled out in Type IIA. This would have been automatically apparent had we written the Type IIA action in string frame, rather than Einstein frame as in \eqref{eq:IIA-action}, where the dilaton does not couple to the RR $p$-forms; our redefinition of $R$ is precisely the relation between string and Einstein frame length scales \cite{ValeixoBento:2025emh}. This emphasises a key point of our analysis: while we can rule out setups that violate the constraints \eqref{eq:constraint-Dd} and \eqref{eq:constraint-dilaton-powers}, we \emph{cannot guarantee} the existence of a minimum for a setup that appears to satisfy them. 

On the other hand a dS$_3$ minimum can be supported by either $F_6$ or $H_7$, with different combinations of the fluxes allowing for both $R$ and $\phi$ to be stabilised. Crucially, we are now allowed to turn on $H_7$, which is required in order to stabilise $\phi$ at a positive minimum. Therefore, one cannot rule out dS$_3$ minima from Type IIA.

\subsection*{\ding{71} $D=9$}

In $D=9$ there is a unique theory with 32 supercharges, which can be regarded as either M-theory on $T^2$ or  as Type IIA or Type IIB on $S^1$ \cite{Bergshoeff:2002nv,Fernandez-Melgarejo:2011nso}. This theory has maximal supersymmetry and only contains the gravity multiplet, which includes three scalars, three vectors, two 2-forms and one 3-form (plus the graviton and fermionic fields),
\begin{equation}
    \{\varphi,\tau=\chi + i e^{-i\phi},A^I,C_1,B_2^a,C_3\} \,,
\end{equation}  
with $I,a = 1,2$. From the M-theory perspective, the vectors $A^I$ correspond to KK vectors in the $T^2$, while the vector $C_1$ arises from dimensionally reducing the M-theory 3-form on the torus; likewise,  the two 2-forms arise from the M-theory 3-form on the two 1-cycles of the torus; as for the scalars, they can be thought of as the complex structure $\tau$ and overall volume modulus $\phi$ of the $T^2$. The bosonic action of maximal supergravity in $D=9$ can then be written as \cite{Fernandez-Melgarejo:2011nso,Bergshoeff:1995as} (setting $2\kappa_9^2=1$), 
\begin{align}
   S_{\rm 9d} &=\, \int \bigg\{
        \star R - \frac12 \ed\varphi\wedge\star\ed\varphi 
        - \frac{\ed\tau\wedge\star\ed\overline{\tau}}{2(\text{Im}\tau)^2} 
        - \frac12 e^{\frac{4}{\sqrt{7}}\varphi}F_2\wedge\star F_2 
        - \frac12 e^{\frac{2}{\sqrt{7}}\varphi}F_4\wedge\star F_4 \nonumber \\
        &\hspace{3em} - \frac12 e^{\frac{3}{\sqrt{7}}\varphi}(\mathcal{M}^{-1})_{IJ}\, F^I\wedge\star F^J 
        - \frac12 e^{-\frac{1}{\sqrt{7}}\varphi}(\mathcal{M}^{-1})_{ab}\, H_3^a\wedge\star H_3^b  \nonumber \\
        &-\frac12 \big[F_4 + \varepsilon_{Ia} A^I\wedge(H_3^a - \frac12\delta^a_{~J}A^J\wedge F_2)\big]\wedge\big(\big[F_4 + \varepsilon_{Ia} A^I\wedge(H_3^a - \frac12\delta^a_{~J}A^J\wedge F_2)\big]\wedge C_1 \nonumber \\
        &\hspace{3em}-\epsilon_{ab}\big[H_3^a - \delta^a_{~I}A^I\wedge F_2\big]\wedge\big[B_2^b -\frac12\delta^b_{~J}A^0\wedge A^J\big]\big)
    \bigg \} \,,
    \label{eq:IIB-action}
\end{align}
where $F^I$ and $F_2$ are the field-strengths of the KK vectors and $C_1$ vector respectively, $H_3^a$ the field-strengths of $B_2^a$, and $F_4$ the field-strength of $C_3$, and $\mathcal{M}$ an $SL(2,\R)$ matrix defined as 
\begin{equation}
    \renewcommand{\arraystretch}{1.1}
    \mathcal{M} = e^\phi \begin{pmatrix}
        |\tau|^2 & \chi \\ 
        \chi     & 1  
    \end{pmatrix} \,.
\end{equation}
Note that $\tau,A^I$ and $B_2^a$ are charged under the perturbative symmetry group\footnote{As usual, at higher energies charged heavy states break this to  the integers $SL(2,\Z)$, which is an exact duality symmetry. } $SL(2,\R)$, transforming as 
\begin{equation}
    \tau\to \frac{a\tau + b}{c\tau + d} \,,\quad
    A^I\to \Omega_J^{\phantom{J}I}A^J \,,\quad 
    B_2^a\to \Omega_b^{\phantom{b}a}B_2^b\,,\quad
    \text{with}\quad\Omega = \begin{pmatrix}
        a & b \\ c & d 
    \end{pmatrix}\in SL(2,\R) \,,
\end{equation}
while $\varphi,C_1,C_3$ are invariant. The magnetic dual $(9-p)$-forms are defined as 
\begin{align}
    F_7 = e^{\frac{4}{\sqrt{7}}\varphi}\star F_2 \,,\quad 
    \tilde{F}_7^I = e^{\frac{3}{\sqrt{7}}\varphi}\star F^I\,,\quad
    H_6^a = e^{-\frac{1}{\sqrt{7}}\varphi}\mathcal{M}^{-1}_{ab}\,\star H_3^a \,,\quad 
    F_5 = e^{\frac{2}{\sqrt{7}}\varphi}\star F_4 \,. 
\end{align}
Therefore, we see that the scalar $\tau$ admits in principle duality freezing. The modulus $\varphi$, however, does not. 

Let us now study moduli stabilization, starting with the volume modulus $R$. In $D=9$, a dS$_4$ minimum requires $p_2 = 5$ flux as demanded by \eqref{eq:constraint-Dd}, which means we must turn on $F_5$ flux, as well as some other flux with $p_1\leq 4$. Turning on both $F_5$ and one of $F_4$, $F_2$ or $H_3^a$ does not work, as it would set off an  axion monodromy effect due to the triple Chern-Simons couplings in the action. The same phenomenon is described in a cleaner setup of 11-dimensional M-theory in Section \ref{sec:charged}; we refer the reader to that section\footnote{And apologize for the forward reference.} for details on the mechanism. The two cases are not disconnected; we can understand the axion monodromy in the $D=9$ case from the M-theory picture by recalling that all of $F_4$, $F_2$ and $H_3^a$ arise from dimensional reduction of the M-theory $G_4$ flux. We are left with a single option for the lower form flux: the 2-forms $F^I$. But in principle, this is enough to stabilize $R$. 

Since the fluxes $F^I$ also transform non-trivially under the duality group, they live in the twisted cohomology group $H^2(\mathcal{M}_k,SL(2,\mathbb{Z}))$. It will be important to take this into account since we will turn on a duality bundle to stabilize $\tau$ in any case. With this caveat, $\tau$ can be stabilised without impairing the three-term structure for the volume modulus, we will not consider it further.

The scalar $\varphi$, corresponding to the M-theory $T^2$, is a different story. It cannot be stabilised by duality bundles, so instead, we must find a three-term structure as we do for the volume modulus. As discussed in Section \ref{s2}, since the Casimir term will not depend on $\varphi$, we must find two flux terms whose $\varphi$ dependences have opposite signs. In this regard, it is fortunate that the only fluxes allowed by the above considerations, $F_5$ and $F^I$, precisely satisfy these conditions, since we have
\begin{equation}V_{F_5}\propto e^{-\frac{2}{\sqrt{7}} \varphi},\quad V_{F^I}\propto e^{\frac{3}{\sqrt{7}} \varphi}.\end{equation}
Therefore, including $F_5$ and $F^I$ fluxes can, in principle, lead to a suitable three-term structure for $R$ and $\varphi$ separately---unfortunately, these ingredients cannot do the trick for both fields at the same time. Schematically, we obtain a scalar potential (before integrating over the volume of the internal space and changing to the 4d Einstein frame)
\begin{equation} 
    V\sim (f_5)^2\,\frac{e^{-\frac{2}{\sqrt{7}} \varphi}}{R^{10}} - \mathcal{C} + \vert f^I\vert^2\,\frac{e^{\frac{3}{\sqrt{7}} \varphi}}{R^4} \,.
\end{equation}
Along the line where $R^4\sim e^{\frac{3}{\sqrt{7}} \varphi}$, the last term is effectively constant, and the potential cannot have a minimum in that direction. This is similar to the IIA case in $D=10$, and illustrates that the more moduli one has, the more fluxes are needed to stabilize them; but here we have run out of options after taking into account our previous constraints. Thus, there are no dS$_4$ maxima in $D=9$ maximal supergravity either. 

Although we did not succeed in finding a dS$_4$, it is worth pointing out an interesting connection between this scenario and the one discussed in \cite{DeLuca:2021pej}. In our $D=9$ construction, we were soon forced to turn on $F^I$ flux. From the M-theory perspective, this is the only flux that does not come from $C_3$, and instead uplifts to pure geometry. It had to be like this; otherwise, we could have decompactified any construction to $D=11$, and applied the no-go established for M-theory there. Since $F^I$ is pure geometry, the ingredients of the $D=9$ construction uplift to M-theory as

\begin{equation} (F_5,\text{Casimir},F^I)\,\xrightarrow{\text{M-theory}} (G_7, \text{Casimir}, \text{Curvature}).\end{equation}
These are precisely the ingredients used in \cite{DeLuca:2021pej} to construct dS$_4$ minima in M-theory. There is however a crucial difference; while \cite{DeLuca:2021pej} uses hyperbolic spaces, the 11D uplift of $F^I$ flux corresponds to replacing the Riemann-flat manifold with an ``almost Riemann-flat manifold'', a nilmanifold \cite{shin1998group,Kaloper:1999yr,Kachru:2002sk} where the $T^2$ is fibered non-trivially over the base.   Nilmanifolds have been studied significantly in the context of string compactifications (see e.g. \cite{Grana:2006kf,Andriot:2015sia,Andriot:2018tmb}); the most general nilmanifold is an iterated sequence of torus bundles over tori \cite{belegradek2002metrics,belegradek2020iterated}.  Since nilmanifolds do have curvature\footnote{In the $D=9$ setup, the contribution of the $F^I$ flux can be made very small, and the metric of the base can be almost flat. It would be interesting to understand the 11d avatar of this mechanism, which implies that in some limits a nilmanifold (which does not admit a Ricci-flat metric) may admit almost-flat metrics where the Casimir energy may still be reliably computed.}, from the point of view of M-theory, we are back in the case of \cite{DeLuca:2021pej}, where no $G_4$ flux is necessary since the role of the lower $p$-form flux is taken by curvature---this prevents the axion monodromies from Chern-Simons terms.
Therefore the $D=9$ case serves as a bridge between Riemann-flat constructions and those involving curvature, and we might have expected a dS$_4$ in this case. As explained above, this expectation fails to materialize since the nilmanifold under consideration contains more moduli (like the $T^2$ volume) that must also be taken into account. In this case they cannot, leading to the lack of dS$_4$ solutions in $D=9$.

Finally, as in the other cases above, dS$_3$ is indeed possible. A dS$_3$ solution requires a flux $4 < p_2 \leq 6$ \eqref{eq:constraint-Dd} which is available, as well as the lower-$p$ fluxes. We also have three moduli (corresponding to the volume and complex structure of $T^2$ from an M-theory perspective), two of which can be fixed via a duality bundle (the complex structure moduli, using the torus $SL(2,\Z)$) and the remaining one (the volume modulus of the $T^2$) might be fixed by fluxes. Casimir-flux dS$_3$ minima on Riemann-flat manifolds are thus possible, in principle, in $D=9$.

\subsection{Theories with all scalars charged under the duality group}
\label{sec:charged}

We now turn to the discussion of theories where duality freezing is most efficient. These all have 32 supercharges, and we will now discuss them by descending dimension.

\subsection*{\ding{71} $D=11$}

There is only one supergravity theory in $D=11$, with 32 supercharges: M-theory. We know from \eqref{eq:constraint-Dd} that in order to get a dS$_4$ minimum we must turn on $G_7$ flux. The solution found in \cite{DeLuca:2021pej} corresponds precisely to this setup, to which negative curvature is added to the $G_7$ flux and Casimir contributions. Since we compactify on Riemann-flat manifolds, we do not have any curvature and must instead resort to a flux with $p_1\leq 5$. In M-theory this can only be $G_4$ flux. At first sight, all seems well, since there are no 11-dimensional moduli to worry about and the fluxes that one needs for the volume potential are available in the theory and can be turned on. There is, however, a problem with this setup. 

The M-theory action contains a Chern-Simons term,
\begin{equation}
    S_{\rm M-theory} \supset \int G_4\wedge G_4\wedge C_3 \,. 
    \label{eq:M-theory-CS}
\end{equation}
The 3-form $C_3$ will give rise, under dimensional reduction, to axions $\theta_i$ coming from the dimensional reduction along three-forms $\omega_3^i$ of the internal space, and a three-form $C_3^{(4d)}$ with all legs in the four-dimensional space, 
\begin{equation}
    C_3 \supset C_3^{(4d)} + \sum_i \theta_i\,\omega^i_3 \,,
\end{equation} 
as well as 2-forms and vector fields that are not important here. Dimensionally reducing the Chern-Simons term \eqref{eq:M-theory-CS} gives rise to a coupling between the three-form and a linear combination of the axions,
\begin{equation}
    S^{(4d)} \supset \int d^4x\,\sqrt{-g_4}\, G_4^{(4d)}\,\left(\sum_i \theta^i n_i \right)\,,
\end{equation}
where the $n_i$ are the coefficients of the expansion of $G_4$ in a basis of 4-forms dual to the $\omega_3^i$,
\begin{equation}
    G_4=\sum_i n_i\,\omega_4^i \,.
\end{equation}
This is precisely a Dvali-Kaloper-Sorbo coupling \cite{Dvali:2005an,Kaloper:2008fb,Kaloper:2008qs}, which induces a potential for the axion, stabilising it at a value that \emph{precisely cancels} any constant value of $G_4^{(4d)}$ in the action. But from the 4d point-of-view, $G_7$ flux is nothing but a background for $G_4^{(4d)}$, so that the much-needed flux for a dS minimum is not available.

Let us discuss in more detail the case where there is a single axion involved. After dimensional reduction, the action for the axion and 3-form is 
\begin{equation}
    S^{(4d)} \supset \int \left(-\frac12\ed\theta\wedge\star_4\ed\theta - \frac12G_4^{(4d)}\wedge\star_4 G_4^{(4d)} + n\, G_4^{(4d)}\theta\right) \,,
\end{equation}   
leading to the equation of motion for the 3-form 
\begin{equation}
    \star_4 G_4^{(4d)} = \star_4\overline{G}_4^{(4d)} + n\,\theta \,, 
\end{equation}
where $\overline{G}_4^{(4d)}$ is some constant 4-form background. Since the 3-form also inherits the gauge symmetry $C_3\to C_3 + \ed\lambda$, it is not dynamical and we can integrate it out using its equation of motion to find the effective action for the axion
\begin{equation}
    S^{(4d)} \supset \int \left(-\frac12\ed\theta\wedge\star_4\ed\theta - (\overline{G}_4^{(4d)} + n\,\theta)^2\right) \,.
\end{equation}  
Turning on the non-trivial background $\overline{G}_4^{(4d)}=(n_7/R^7) dV$ in terms of the $G_7$ flux $n_7$, the above leads to a modified flux potential (after rescaling $\theta$ by an $R$-dependent factor)
\begin{equation}
    V_{G_7,\text{Monodromy}}\sim \frac{(n_7 + n\,\theta)^2}{R^7}\,,
\end{equation}
and extremizing $\theta$ destroys the three-term structure for $R$ \cite{Dvali:2005an,Kaloper:2008fb,Kaloper:2008qs}. 
The upshot is that, in the presence of these Chern-Simons couplings, any $G_7$ flux contribution is washed out by the axion expectation value, along with the dS$_4$ minimum for $R$. 

A dS$_3$ solution from M-theory still requires $G_4$ and $G_7$ fluxes to be turned on, but now the compact space is 8-dimensional and the problematic axionic coupling from the Chern-Simons term \eqref{eq:M-theory-CS} may be avoided. Since there are no further obstructions, we conclude that a Casimir-flux dS$_3$ is, in principle, possible in M-theory on Riemann-flat manifolds.

\subsection*{\ding{71} $D=10$}

In $D=10$ there is one theory with 32 supercharges that does not require fluxes to stabilise the dilaton, Type IIB supergravity. However, constraint \eqref{eq:constraint-Dd} tells us that, to stabilize the volume, a dS$_4$ solution requires a 6-form flux, which is absent in Type IIB. We can therefore not construct a Casimir-flux dS$_4$ in Type IIB with Riemann-flat manifolds. 

A dS$_3$ minimum requires, in order to stabilize the volume, a flux with $5<p_2\leq 7$ \eqref{eq:constraint-Dd}, whose role can be played in Type IIB by either $H_7$ or $F_7$, while a combination of $F_1,F_3,H_3$ fluxes can play the role of the $p_1$-flux. So from the point of view of the volume modulus, a Type IIB dS$_3$ minimum is not automatically excluded. Since Type IIB has an $SL(2,\Z)$ duality group, we can consider   duality freeze-out to stabilise the dilaton. 
Consider the Type IIB action, 
\begin{align}
    2\kappa_{10}^2\,S_{\rm IIB} =\,& \int d^{10}x \left(R - \frac12(\partial\phi)^2 - \frac{e^{-\phi}}{2}|H_3|^2 - \frac{e^{2\phi}}{2}|F_1|^2 - \frac{e^\phi}{2}|F_3|^2 - \frac{1}{4}|F_5|^2\right) \nonumber \\
    & -\frac12\int C_4\wedge H_3\wedge F_3 \,. 
\end{align}
The complex field $\tau = C_0 + i\,e^{-\phi}$ (with $F_1 = \ed C_0$) and the field-strengths $F_3$ and $H_3$ transform under $SL(2,\Z)$ as 
\begin{align}
    \tau\to\frac{a\tau + b}{c\tau + d}\,,
    && \begin{pmatrix}
        F_2 \\ H_2
    \end{pmatrix} \to \begin{pmatrix}
        a & b \\
        c & d
    \end{pmatrix}\begin{pmatrix}
        F_3 \\ H_3
    \end{pmatrix}\,,
    &&\text{with}\quad\begin{pmatrix}
        a & b \\ 
        c & d
    \end{pmatrix}\in SL(2,\Z) \,.
\end{align} 
In practice, ``turning on a non-trivial duality bundle'' corresponds to choosing boundary conditions for the fields on a given RFM such that each field goes to its $SL(2,\Z)$ image as one goes around a non-contractible cycle of the RFM. For example, requiring the fields to match their image under the S-duality transformation 
\begin{align}
    \tau\to -\frac{1}{\tau} \,, 
    \quad\begin{pmatrix}
        F_3 \\ H_3
    \end{pmatrix} \to \begin{pmatrix}
        -H_3 \\ F_3
    \end{pmatrix} \,,
    \quad\begin{pmatrix}
        F_7 \\ H_7
    \end{pmatrix} \to \begin{pmatrix}
        -H_7 \\ F_7
    \end{pmatrix} \,,
\end{align}
upon going around some cycle of the RFM, would freeze out $C_0 = 0,\,e^\phi = 1,\,H_3=F_3=H_7=F_7=0$. In other words, we are choosing boundary conditions that are consistent in the higher-dimensional theory but forbid zero modes for these fields. 

While this automatically freezes the scalars, including the dilaton\footnote{One may be worried about the fact that this happens in the strong coupling regime $e^{\phi}=1$. However, the low-energy effective action is fully fixed by supersymmetry, even at strong coupling. One may compare this to the situation of M-theory,  where there is also no small coupling,  yet compactifications on manifolds much larger than the eleven-dimensional Planck length are under control. This is precisely because the M-theory action (at the two-derivative level) is fully fixed by supersymmetry.}, it also forces those fluxes charged under $SL(2,\mathbb{Z})$ to live in appropriately twisted cohomology groups. In particular, this applies to the $(H_7,F_7)$ required to stabilize the volume. However, it is a general fact that, for $\mathcal{M}_k$ orientable, and a duality group $\Gamma$ leaving no invariant $k$-forms, 
\begin{equation} H^k(\mathcal{M}_k,\Gamma)\otimes \mathbb{Q}=0,\end{equation}
i.e. the free part of the twisted cohomology of top degree is trivial. This happens because untwisted cohomology is one-dimensional, and if $\Gamma$ has no fixed points itself, there can be none at all. We can thus not use duality freeze-out to fix the dilaton in this case without destroying $R$ stabilization.

However, since we are now allowed to turn on 7-form fluxes, it is in principle possible to stabilise the dilaton, e.g. by turning on the NS-NS dual pair $H_3/H_7$, just as in the Type IIA case\footnote{The discussion regarding the RR flux terms in Type IIA applies equally to this Type IIB setup; RR fluxes cannot be relied on to stabilise the dilaton.}. This flux choice does not turn on any axion monodromy effects, since the Chern-Simons terms always involve one NS and two RR fields.  Note also that the axion $C_0$ has a shift symmetry and is in no danger of developing a runaway, at least at weak coupling. This means we cannot rule out a Type IIB Casimir-flux dS$_3$ minimum with Riemann-flat manifolds.

\subsection*{\ding{71} $D=8$}

In $D=8$, dS$_4$ minima are ruled out by condition \eqref{eq:constraint-Dd}. 
In order to find a dS$_3$ minimum, we need a flux with $p_2 = 5$, which does exist in 8d supergravity with 32 supercharges. The duality group is now $SL(2,\Z)\times SL(3,\Z)$, which we could use to fix the remaining moduli through a duality freeze-out---there are two scalars transforming under $SL(2,\Z)$ and five scalars transforming under $SL(3,\Z)$. However, the available 5-forms (that are magnetic duals of 3-forms) transform in the $(\mathbf{1},\mathbf{3})$ representation of the duality group (see Table \ref{tb:duality-groups}), so that any duality bundle that can freeze all scalars will also necessarily project out all 5-forms. Therefore, just like in the IIB example above, we cannot use duality freeze-out to get rid of all the scalars. 

Let us be more explicit. 
The field content of 8d supergravity with 32 supercharges is  \cite{Alonso-Alberca:2000wkg,LassoAndino:2016lwl,Sezgin:2023hkc}
$$\{G_{MN},C_3,B_2^m,A^{i,m},\phi^I,\varphi,C_0\}$$
where $m=1,...,3$, $i=1,2$, $I=1,...,5$. The 2-forms and scalar $\phi^I$ are charged under the $SL(3,\Z)$ part of the duality group; the complex scalar $\tau = C_0 + ie^\varphi$ transforms under the $SL(2,\Z)$ part of the duality group; and the vectors $A^{i,m}$ transform under both $SL(3,\Z)$ and $SL(2,\Z)$. This implies that $F_4$ flux is duality-invariant; $H_3^m/H_5^m$ fluxes transform in the $\mathbf{3}$ representation of $SL(3,\Z)$; $F_2^{i,m}/F_6^{i,m}$ fluxes transform under the $(\mathbf{2},\mathbf{3})$ representation of the duality group; and $F_1/F_6$ flux transforms under $SL(2,\Z)$.

Stabilising the volume modulus requires $H_5^m$ flux---of which there are 3---and at least one of $F_1/F_2^{i,m}/H_3^m$---of which there are 10 in total. Since $H_5^m$ transforms under $SL(3,\Z)$, we cannot use duality freeze-out to fix the five scalars $\phi^I$ without losing the 5-form flux needed for the volume modulus. Thus, these must be stabilised through fluxes. If we are to fix  $\tau$ with an $SL(2,\Z)$ duality bundle, the fluxes $F_2^{i,m}$ must live in twisted cohomology; as in other examples above, this is not an obstacle in principle. Therefore, we will assume that $\tau$ is fixed by duality freeze-out and all of $H_5^m,H_3^m,F_4, F_{2}^{i,m}$ fluxes are available to us.

The action is given by \cite{Alonso-Alberca:2000wkg}
\begin{align}
    S = \int\Bigg\{&
        ... -\frac{1}{2\cdot 2!}\mathcal{W}_{ij}\mathcal{M}_{mn}F_2^{i,m}\wedge\star F_2^{j,n}
        + \frac{1}{2\cdot 3!}(\mathcal{M}^{-1})_{mn}H_3^m\wedge\star H_3^n 
        -\frac{1}{2\cdot 4!}e^{-\varphi}F_4\wedge\star F_4
    \nonumber \\
    &-\bigg[F_4\wedge F_4\wedge C_0
        -8F_4\wedge H_3^m\wedge A^{2,m}
        +12 F_4\wedge(F_2^{2,m} + C_0\,F_2^{1,m})\wedge B_2^m  \\ 
    &\quad    -8F_4\wedge F_1\wedge C_3 
        -\epsilon_{mnp}H_3^m\wedge H_3^n\wedge B_2^p 
        -16H_3^m\wedge(F_2^{2,m} + C_0\,F_2^{1,m})\wedge C_3
    \bigg]
    \Bigg \}\nonumber \,.
\end{align}
We see that the 4-form flux appears in many Chern-Simons terms that could cause axion monodromies as the one we found in M-theory.  As we see in the action above, $F_4$ flux only couples to the volume and the scalar $\varphi$, which is part of $\tau$. Since we assume that $\tau$ is fixed through duality bundles, this flux does not help at all to stabilise the moduli, so we can switch it off and avoid these potentially dangerous Chern-Simons terms. The five scalars that cannot be stabilised with duality bundles are encoded in the matrix $\mathcal{M}_{mn}$, which means that they can get a potential from $F_2^{i,m}/H_3^m/H_5^m$ fluxes. The potential for the scalars becomes 
\begin{equation}
    V\propto \frac{\mathcal{W}_{ij}\mathcal{M}_{mn}n_2^{i,m}n_2^{j,n}}{R^4} 
    + \frac{(\mathcal{M}^{-1})_{mn}n_3^m n_3^n}{R^6}  
    -\frac{\mathcal{C}}{R^{8}}
    + \frac{\mathcal{M}_{mn}n_5^m n_5^n}{R^{10}} \,.
\label{fff}\end{equation}
It should be kept in mind that not all scalars in $\mathcal{M}_{mn}$ need to be stabilised for a $dS$ minimum to exist: some of these are compact scalars (axions) \cite{Cremmer:1997ct}, whose value even if unstabilised will not affect the properties of the solution as they can never lead to a runaway. The five scalars charged under $SL(3,\mathbb{Z})$ admit a physical interpretation as the shape moduli of the M-theory $T^3$; in a certain parametrization, three of these can be taken to be compact and two remain non-compact. The three compact scalars provide three additional 1-form fluxes that can in principle be added to \eq{fff} (they come from the kinetic term of the $\mathcal{M}_{mn}$) and which depend on different combinations of the non-compact scalars, bringing the total count to seven flux terms and three non-compact scalars including the volume modulus $R$.
In this case, unlike the ones we encountered before, it is not easy to find a field redefinition that clearly shows that the requisite three-term structure is absent for some particular combination of scalars, as we did in Section \ref{sec:uncharged-32}. Although it may exist, finding it would require a more detailed analysis that is beyond the scope of this paper. As a result, we give up and declare that dS$_3$ minima are not excluded in principle, though we emphasize that a more thorough analysis might very well end up excluding this case as well.

\subsection*{\ding{71} $D=7$}

In $D=7$, dS$_4$ minima are ruled out by condition \eqref{eq:constraint-Dd}. For a dS$_3$ minimum, we need a flux with $p_2 = 4$, but all available 4-forms are charged under the theory's duality group---they transform in the $\mathbf{5}$ representation of $SL(5,\Z)$ (Table \ref{tb:duality-groups}). In a compactification to $d=3$, the 4-forms are top forms---since there is no twisted cohomology in top degree, any duality bundle that allows to freeze all moduli will necessarily project out all these 4-forms. Therefore, everything must be stabilised via fluxes in this case. The analysis is in fact very similar to the $D=8$ case above; the field content of 7d supergravity with 32 supercharges is \cite{Sezgin:1982gi}
$$\{G_{MN},B_2^m,A^{mn},\phi^I\}$$
with $B_2^m$ transforming in the $\mathbf{5}$ of $SL(5,\Z)$, the vectors $A^{mn}$ transforming in the $\mathbf{10}$ of $SL(5,\Z)$, and scalars $\phi^I$ transforming as the shape moduli of a fictitious $T^5$. Just like in $D=8$, this contains 14 moduli, split as 10 compact and 4 non-compact scalars \cite{Cremmer:1997ct}. 
In total, we can switch on $H_3^m/H_4^m$, $F_2^{mn}$, and 1-form fluxes coming from the periodic scalars; either $H_3^m$, $F_2^{mn}$ or the 1-form fluxes could play the role of the $p_1$-form flux needed for the stabilisation of the volume modulus; we now also need to stabilize the four non-compact scalars. We are again in a complicated case like $D=8$ in the previous subsection, and therefore we conclude that dS$_3$ is in principle possible here---though we would not be surprised if a motivated reader could rule these out too.

\subsection{Summary}
A summary of the results obtained throughout this section can be found in Table \ref{tab:theories-32supercharges} for theories with 32 supercharges; and Table \ref{tab:theories-16supercharges} for theories with 16 supercharges. The relevant duality groups as well as the number of fields and the representations they are charged under are contained in Table \ref{tb:duality-groups}. Overall, we find that dS$_4$ is not possible in the context of RFM flux compactifications with Casimir energy, while dS$_3$ cannot be definitively ruled out in any theory.

{\renewcommand{\arraystretch}{1.5}
\begin{table}[!htb]
    \centering
    \normalsize
    \begin{tabular}{|c|c|c|c|l|}
         \hline 
         \rowcolor{gray!10} $d$ & $D$ & $p_2$ & dS$_d$ & Comment \\
         \hline 
         \multirow{4}{*}{$4$} & $11$ &  $6,7$ & \xmark & There is $G_7$ and no moduli, but axion monodromy cancels $G_7$ \\ \cline{2-5}
                                & \multirow{2}{*}{$10$} & \multirow{2}{*}{$6$} & \xmark & IIA: $F_6$ exists, but $\phi$ cannot be stabilised \\ \cline{4-5}
                                & & & \xmark & IIB: No $F_6$, although $\tau$ could be fixed with $SL(2,\Z)$ bundle \\ \cline{2-5}
                                & 9 & 5 & \xmark  & $F_5/F^I$ fluxes could stabilise $R$, but $\varphi$ is not stabilised \\ \hline
        \multirow{9}{*}{$3$}   & $11$ & $6,7,8$ & \emptycell & $G_7$ exists, no moduli \\ \cline{2-5}
                                & \multirow{3}{*}{$10$} & \multirow{3}{*}{$6,7$} & \emptycell & IIA: $F_6$ exists, as well as $H_3/H_7$ to stabilise $\varphi$ \\ \cline{4-5}
                                & & & \multirow{2}{*}{\emptycell} & IIB: $F_7$ exists; if $\tau$ is fixed with $SL(2,\Z)$ bundle, $F_7$ is projected \\ 
                                & & & & out. Fluxes needed to stabilise $\phi$, but possible in principle. \\ \cline{2-5}
                                & $9$  & $5,6$ & \emptycell & $F_5 / F_6$ exist, as well as $H_3/H_7$ to stabilise $\phi$ \\ \cline{2-5}
                                & \multirow{2}{*}{$8$}  & \multirow{2}{*}{$5$} & \multirow{2}{*}{\emptycell } & $F_5$ exists; if scalars are fixed with $SL(3,\Z)$ bundle, $F_5$ is projected \\ 
                                & & & & out. Fluxes needed to stabilise scalars, but possible in principle.  \\ \cline{2-5}
                                & \multirow{2}{*}{$7$}  & \multirow{2}{*}{$4$} & \multirow{2}{*}{\emptycell} & $F_4$ exists; if scalars are fixed with $SL(5,\Z)$ bundle, $F_4$ is projected  \\
                                          & & & & out. Fluxes needed to stabilise scalars, but possible in principle. \\ \hline
    \end{tabular}
    \caption{Theories in $D$ dimensions with 32 supercharges that could give rise to a $d$-dimensional dS with a potential generated by fluxes and Casimir energy; we include the type of $p_2$-flux required to stabilise the volume modulus $R$ by \eqref{eq:constraint-Dd}, as well as a cross for all cases in which dS$_d$ minima can be ruled out. Note that no dS$_4$ minima survive, but dS$_3$ minima cannot be ruled out, in principle.}
    \label{tab:theories-32supercharges}
\end{table}}

{\renewcommand{\arraystretch}{1.5}
\begin{table}[!htb]
    \centering
    \begin{tabular}{|c|c|c|c|c|c|}
        \hline 
        \rowcolor{gray!10} $D$ & $p$-forms & $d$ & $p_2$ & dS$_d$ & Comment \\ \hline
        \multirow{2}{*}{10} & \multirow{2}{*}{$F_2,H_3,H_7,F_8$} & 4 & 5,6 & \xmark & No flux can be $p_2$ \\ \cline{3-6} 
        & & 3 & 6,7 & \emptycell & Needs $H_7$ and $\phi$ stabilised via fluxes \\ \hline  
        \multirow{2}{*}{9} & \multirow{2}{*}{$F_2,H_3,H_6,F_7$} & 4 & 5 & \xmark & No flux can be $p_2$ \\ \cline{3-6} 
        & & 3 & 5,6 & \emptycell & Needs $H_6$ and $\phi$ stabilised via fluxes \\ \hline  
        \multirow{2}{*}{8} & \multirow{2}{*}{$F_2,H_3,H_5,F_6$} & 4 & \emptycell & \xmark & No value of $p_2$ allowed \\ \cline{3-6} 
        & & 3 & 5 & \emptycell & Needs $H_5$ and $\phi$ stabilised via fluxes \\ \hline  
        \multirow{2}{*}{7} & \multirow{2}{*}{$F_2,H_3,H_4,F_5$} & 4 & \emptycell & \xmark & No value of $p_2$ allowed \\ \cline{3-6} 
        & & 3 & 4 & \emptycell & Needs $H_4$ and $\phi$ stabilised via fluxes \\ \hline         
    \end{tabular}
    \caption{Theories in $D$ dimensions with 16 supercharges, together with the corresponding available fluxes, that could give rise to a $d$-dimensional dS with a potential generated by fluxes and Casimir energy; we include the type of $p_2$-flux required by \eqref{eq:constraint-Dd}, as well as a cross for all cases in which dS$_d$ can be ruled out. Note that all dS$_4$ minima are ruled out for theories with 16 supercharges and that dS$_3$ minima require at least $H_{D-3}$ flux to be turned on. In all cases there is a dilaton that must be stabilised by fluxes.}
    \label{tab:theories-16supercharges}
\end{table}}

{\renewcommand{\arraystretch}{1.5}
\begin{table}[!htb]
    \centering
    \scalebox{0.84}{\begin{tabular}{|c|c|c||c||c|c|c||c|}
        \hline 
        \rowcolor{gray!10} $N_Q$ & $D$ & Duality group & Scalars & $p=1$ & $p=2$ & $p=3$ & Comment  \\ \hline
        \multirow{5}{*}{32} 
        & 11 & \emptycell & \emptycell & \emptycell & \emptycell & \emptycell & M-theory \\ \cline{2-7} 
        & \multirow{2}{*}{10} & \emptycell & \emptycell & \emptycell & \emptycell & \emptycell & Type IIA \\ \cline{3-7}
        & & $SL(2,\Z)$ & $2$ & \emptycell & $\mathbf{2}$ & \emptycell & Type IIB \\ \cline{2-7} 
        & 9 & $SL(2,\Z)$ & $3$ & $\mathbf{1}\oplus\mathbf{2}$ & $\mathbf{2}$ & $\mathbf{1}$ & M-theory on $T^2$ \\ \cline{2-7} 
        & 8 & $SL(2,\Z)\times SL(3,\Z)$ &  $5$ & $(\mathbf{2},\mathbf{3})$ & $(\mathbf{1},\mathbf{3})$ & $(\mathbf{2},\mathbf{1})$ & M-theory on $T^3$ \\ \cline{2-7} 
        & 7 & $SL(5,\Z)$ & $14$ & $\mathbf{10}$ & $\mathbf{5}$ & \emptycell & M-theory on $T^4$ \\ \hline
        \multirow{4}{*}{16}  
        & 10 & \emptycell & \emptycell & \emptycell & \emptycell & \emptycell & Heterotic/Type I \\ \cline{2-7} 
        & 9 & $SO(1,n_V;\Z)$ & $n_V+1$ & $\mathbf{18}$ & $\mathbf{1}$ & \emptycell & Heterotic on $S^1$ \\ \cline{2-7} 
        & 8 & $SO(2,n_V;\Z)$ & $2n_V + 1$ & $(\mathbf{1},\mathbf{20})$ & $\mathbf{1}$ & \emptycell & Heterotic on $T^2$ \\ \cline{2-7} 
        & 7 & $SO(3,n_V;\Z)$ & $3n_V + 1$ & $(\mathbf{1},\mathbf{22})$ & $\mathbf{1}$ & \emptycell & Heterotic on $T^3$ \\ \hline 
    \end{tabular}}
    \caption{Duality groups for theories with 32 and 16 supercharges in $D=7,...,11$ \cite{Narain:1985jj,Narain:1986am,Obers:1998fb,Samtleben:2008pe,Braeger:2025kra}. We include the number of scalars transforming non-linearly under the duality group (see text for details in each theory), the representations of the duality group under which $p$-form fields transform, as well as an embedding of each theory in a known string/M-theory.}
    \label{tb:duality-groups}
\end{table}
}

\newpage
\section{dS searches and the semidefinite linear bootstrap}
\label{sec:comments-dS3}

As we have seen in the previous sections, it is harder to rule out dS$_3$ minima on general grounds. In principle, several $D$-dimensional theories with 16 or 32 supercharges allow for a Casimir-flux de Sitter minimum on Riemann-flat manifolds in $d=3$. However, the conditions we imposed so far are only necessary, rather than sufficient; therefore none of the surviving setups is guaranteed to have a dS$_3$ minimum. 

To find out whether a dS$_3$ minimum exists in any given setup requires a detailed analysis which is beyond the scope of this work. If one is lucky, a potential with the requisite three-term structure will eventually be achieved for all unstabilised moduli. Even then, the existence of a dS$_3$ minimum requires a delicate balance between the coefficients of these terms. Building on  a key idea introduced in \cite{Grimm:2019ixq}---that searches for vacua in flux compactification can be attacked with tools of linear optimisation---,  we will show that the problem of finding appropriate tunings in the coefficients to obtain a dS$_3$  can be recast as a semidefinite optimisation problem, of the kind familiar from the CFT bootstrap \cite{Simmons-Duffin:2015qma,Chester:2019wfx,Simmons-Duffin:2016gjk}. 

To show this, let us go back to the problem of finding de Sitter. As discussed in previous sections, the potential for the moduli takes the generic form
\begin{equation}
    \frac{V(R,\phi^I)}{(M_d)^d} = \frac{1}{R^{\frac{d}{d-2}k}}\left(\sum_{p} \frac{f_p(\phi^I)}{R^{2p-k}} - \frac{\mathcal{C}(\phi^I)}{R^d}\right) \,, 
\end{equation}
where $R$ again denotes the volume modulus, expressed in $d$-dimensional Planck units, and $\phi^I$ other moduli of the compactification from $D$-dimensions down to a $d$-dimensional EFT that have not been duality-frozen. The reduction has taken place on a $(k=D-d)$--dimensional RFM, and we have also included an overall factor coming from rescaling to Einstein frame.  The potential depends on flux terms encoded in the functions $f_p(\phi^I)$; each of these is a quadratic function on the integer-quantised $p$-form fluxes $n_p^\alpha$, 
\begin{equation} 
    f_{p}(\phi^I)=\sum_{\alpha,\beta} \mathcal{G}^{(p)}_{\alpha \beta} n_p^\alpha n^\beta_p \,,
\end{equation}
where $ \mathcal{G}^{(p)}_{\alpha \beta}$ is the appropriate Hodge norm for the $p$-forms involved (which may or may not live in twisted cohomology). Importantly, since the kinetic term of these $p$-form fields are all positive definite, we have $f_p\geq 0$, with equality achieved only for vanishing fluxes. The potential also depends on the Casimir term $\mathcal{C}(\phi^I)$, which can have a complicated dependence on the $\phi^I$ in general. The explicit form of $\mathcal{C}(\phi^I)$ can be computed for any given RFM, once the $D$-dimensional theory and spin structure are fixed \cite{ValeixoBento:2025yhz,DallAgata:2025jii}. For a dS minimum, it is crucial that the Casimir energy contribution to the potential is negative, so we will assume $\mathcal{C}(\phi^I)>0$. We can now define the positive quantities  
\begin{align}
    V_{p} = \frac{f_p(\phi^I)}{R^{2p-k}} \,,
    \quad \Vcas = \frac{\mathcal{C}(\phi^I)}{R^d} \,.
    \label{eq:VpVcas-def}
\end{align}

In terms of these, the conditions for a Sitter critical point take a simple form,
\begin{subequations}
    \begin{align}
    \partial_R V &\propto -\sum_{p} 2(k + (d-2)p)V_{p} + d(d+k-2)\Vcas = 0 \,, \\   
    \partial_I V &\propto \sum_{p} (\partial_I\log f_{p})\,V_{p} - (\partial_I\,\mathcal{C})\,\Vcas = 0 \,, \\
    0 &= \sum_p V_p - \Vcas - V_0 \,,
\end{align}
\label{eq:sdp-equalities}
\end{subequations}
where we have also introduced the value of the potential at the critical point, $V_0>0$. Crucially, since the dependence on $R$ on each term is polynomial, the first and second derivatives are linear functions of $V_p,V_0$ and $\Vcas$. 
The problem then can be rewritten as a matrix equation
\begin{equation}
    \mathbf{A}\,\vec{v} = 0 \,,
    \label{sys0}
\end{equation}
with  $\vec{v}^{\,T} = (V_{1},\ldots, V_q, \Vcas,V_0) > 0$ a $(q+2)$-dimensional vector of positive or zero entries, where $q$ is the number of non-zero flux terms turned on, and $\mathbf{A}$ a $(q+2)\times (N+2)$ matrix, where $N$ is the total number of moduli excluding $R$, which takes the form
\begin{equation}
    \renewcommand{\arraystretch}{1.1}
    \mathbf{A} = \begin{pmatrix}
        -2(k + (d-2)p_1) & \cdots & -2(k + (d-2)p_q) & d(d+k-2)  & 0 \\
        (\partial_1\log f_{p_1})  & \cdots & (\partial_1\log f_{p_q}) & -(\partial_1\,\mathcal{C})       & 0 \\ 
        \vdots & \ddots & \vdots & \vdots & 0 \\
        (\partial_N\log f_{p_1}) & \cdots & (\partial_N\log f_{p_q}) & -(\partial_N\,\mathcal{C})  & 0 \\ 
        1             & 1  & 1 & -1         & -1 
    \end{pmatrix} \,.
\end{equation}

According to Stiemke's theorem \cite{stiemke1915positive}, either the homogeneous system of equations \eqref{sys0} has a solution with $\vec{v}>0$, or the system of inequalities 
 \begin{equation}
    \mathbf{A}^T \,\vec{y}>0
    \label{stinky}
\end{equation}
has a solution. The latter is often more tractable, and this  was exploited in \cite{Grimm:2019ixq} to generalise some no-go theorems about de Sitter. Note that there will be as many inequalities as terms in the potential, plus one that simply encodes de sign of the potential at the critical point; this means one could also study AdS solutions by simply flipping the sign of $V_0$. Setting a given term $\log f_{p_{i}}$ to zero (e.g. turning off certain fluxes) is equivalent to removing an inequality from the system---since the number of moduli remains constant, the remaining inequalities can only be easier to satisfy. In other words, less terms in the potential only make it harder to find a solution. 

As a simple but key example, consider removing the inequality coming from the Casimir term altogether. The system of inequalities \eq{stinky} reduces to 
\begin{equation}
    -2(k+(d-2)p_i)\,y_R + \sum_I (\partial_{I}\log f_{p_i})\,y_I + y_0> 0 \,,
    \label{eq:ineq-fluxes}
    \quad \forall p_i \,,
    \quad y_0<0 \,,
\end{equation}
which \emph{can} be satisfied by taking e.g. $y_I=0$, $y_0=y_R=-1$.
In effect, we simply recover the fact that a potential that only contains flux terms has no critical point for $R$. We can also show that a dS minimum requires at least one flux term with $2p>D$, as in our constraint \eqref{eq:constraint-Dd}. To see this, we add back the Casimir term in the potential and focus on the corresponding inequality
\begin{equation}
    d(d+k-2)y_R - \sum_I (\partial_I\mathcal{C})\,y_I - y_0 > 0 \,.  
    \label{eq:ineq-Casimir}
\end{equation}
First note that the simple solution $y_I=0$, $y_0=y_R=-1$ no longer solves the system of inequalities. Adding \eqref{eq:ineq-Casimir} and \eqref{eq:ineq-fluxes}, we find
\begin{equation}
    (d-2)(D-2p_i)y_R + \sum_I[(\partial_I\log f_{p_i}) - (\partial_I\mathcal{C})]y_I > 0 \,, \quad\forall p_i \,,
\end{equation}
with $D = d + k$ the dimension of the higher-dimensional theory. Taking as before $y_I = 0$, this set of inequalities is satisfied for any $y_R < 0$ \emph{if} $D<2p_i$ for all fluxes $p_i$. This automatically satisfies \eqref{eq:ineq-fluxes} and it will satisfy \eqref{eq:ineq-Casimir} if we choose $y_0 < d(D-2)y_R < 0$. Thus a dS minimum requires at least one flux to satisfy $2p>D$, in accordance with our constraint \eqref{eq:constraint-Dd}.

So far we have only imposed the existence of a dS critical point. For this solution to be a \emph{minimum} of the potential we also need the Hessian $\mathcal{H}_{IJ} = \partial_I\partial_J V$ to be positive-definite, or positive-semidefinite (which we denote as $\mathcal{H}\succeq0$) if we allow for flat directions in our approximation. The Hessian depends linearly on the $V_p$ and $\Vcas$, and is independent of $V_0$. We can therefore write
\begin{equation}
    \mathcal{H}_{IJ}= \sum_i h^{i}_{IJ} \vec{v}_i \succeq 0 \,,
    \label{eq:sdp-inequality}
\end{equation}
where the matrices $h_{IJ}^i$ can be computed explicitly as we did for the gradient above; we omit the precise expressions since they would only add clutter.

To summarize the above: Finding a dS minimum corresponds to finding a non-vanishing vector $\vec{v}^{\,T}=(V_p,\Vcas,V_0)$, where all components are bigger than or equal to zero, subject to the equalities \eq{eq:sdp-equalities} and to the inequality \eq{eq:sdp-inequality}. We can produce a more homogeneous presentation of the problem by rephrasing the equalities in terms of positive-semidefinite conditions for matrices. For instance, demanding $\vec{v}>0$ and \eq{eq:sdp-equalities} amounts to 
\begin{equation} 
    M_a\succeq0 \,, \quad a=1,2,3 \,,
\end{equation}
where 
\begin{equation} 
    M_1\equiv\text{diag}(\vec{v}_1,\ldots \vec{v}_{q+2})\,, 
    \quad M_2\equiv\text{diag}((\mathbf{A}\, \vec{v})_1,\ldots (\mathbf{A}\, \vec{v})_{N+2})\,, 
    \quad M_3\equiv-M_2 \,.
\end{equation}
If we add $M_4=\mathcal{H}_{IJ}$, then the problem of finding a dS minimum can be recast succintly as
\begin{equation} 
    \text{Find}\, \,\vec{v}\in \mathbb{R}^{q+2}\,\text{such that}\quad M_a(\phi^I,R)\succeq0\quad  \text{for}\, a=1,2,3,4 \,.
    \label{diff}
\end{equation}
This is a semidefinite programming problem. More concretely, it is exactly a ``Matrix Program'' of the kind introduced in \cite{Simmons-Duffin:2015qma}. That paper introduces a numerical solver adapted to the kind of semidefinite programs stemming from CFT bootstrap; we now see that the search for de Sitter yields exactly the same problem! In the simplest CFT bootstrap problems, one is looking for values of the conformal dimensions $\Delta_i$ such that there are choices of three-point function coefficients $\lambda_{ijk}$ for which the crossing symmetry equations for a number of correlation functions are satisfied. The map between CFT bootstrap data and our dS search is 
\begin{align} 
    \Delta_i\quad&\leftrightarrow\quad \phi^I, \, R \nonumber \\
    \lambda_{ijk}\quad&\leftrightarrow\quad \vec{v}=(V_p,\Vcas,V_0) \nonumber \\ 
    \text{Crossing symmetry}\quad&\leftrightarrow\quad \text{dS critical point conditions} \label{rert}  \\ 
    \text{A particular CFT}\quad&\leftrightarrow\quad \text{A particular RFM} \,. \nonumber 
\end{align}
On top of this, the dS search has three features not present in the traditional CFT bootstrap: 
\begin{itemize}
    \item  CFT boostrap problems are what \cite{Simmons-Duffin:2015qma} calls polynomial matrix programs (PMP), where the entries of the matrices above are all polynomial.  This is because derivatives of conformal blocks can be recast as polynomials of the conformal cross-ratios times positive functions. This fact is key for the efficient numerical implementation achieved in \cite{Simmons-Duffin:2015qma}. Instead, dS search matrices are not polynomial in the moduli,  due to the often complicated dependence of Casimir energy and fluxes. However, one can simply Taylor-expand these around reference points, turning the problem into a PMP. Moreover, if one is willing to pay the computational price, or if the number of moduli is low enough,  it  is possible to perform a direct scan for values of $(\phi^I, R)$ that harbor a de Sitter minimum. This is what we will do in the example below. It is worth exploring in some detail whether the semidefinite program for dS searches can be turned into a PMP by some trick or redefinition, but we will not attempt this here.
    \item Valid dS minima require flux quantisation, which manifests itself in the fact that the vector $\vec{v}$ can only take discrete values. In practice this will not matter much for the searches, as one can always ignore quantisation at first to find the allowed $\vec{v}$ region and only later determine whether there any properly quantised points lying in it. We hope that the map \eq{rert} can be helpful in leveraging semidefinite programming techniques for the search of dS vacua, particularly in situations with many moduli where the existing efficient numerical techniques may make a difference.
    \item The third novel feature in dS searches is that the parameter range is restricted, since we must demand that any solution is in a reliable region of moduli space. The two-derivative action and Casimir potential we use have been derived for large-volume compactifications where $R\gg1$; we are forced to discard any solution where the radius is not large enough\footnote{In fact, the condition one must impose is that every cycle in the RFM is large enough in the appropriate units; depending on the RFM there may be some hierarchy between $R$ and the size of the smallest cycles (e.g. in the dS$_5$ saddle of \cite{ValeixoBento:2025yhz} we have moduli $\{R,x\}$ and shortest cycles $\frac18 R\,x^{-2/3}\,, \sqrt{2}R\,x^{1/3}$ that must be significantly larger than one). Likewise, proper flux quantisation follows from identifying the correct cycles in cohomology for the chosen RFM (e.g. in the setup of \cite{ValeixoBento:2025yhz} the $\Z_8$ quotient decreases the $n_3$ flux quantum by a factor of $8$).}. This can be conveniently encoded in terms of an \emph{objective} function. As explained in \cite{Simmons-Duffin:2015qma}, the most general semidefinite program involves optimizing a linear function $\vec{b}\cdot\vec{v}$ over a set of $\vec{v}\,'$s satisfying the constraints derived above. In CFT bootstrap one is only interested in feasibility, so $\vec{b}$ is set to zero. We will instead use the linear objective function to encode the consistency condition $R\gg1$. To do this, we will use the fact that the flux and Casimir components of $\vec{v}$ have a well-defined scaling with $R$; for instance, $\Vcas=\mathcal{C}\,R^{-d}$, so that for fixed values of the moduli $(\phi^I,R)$ (and hence fixed $\mathcal{C}$), maximizing $R$ is equivalent to minimizing $\Vcas$. Therefore, we will complement the semidefinite program \eq{diff} by the additional requirement that we wish to minimize $\Vcas$. Other choices are certainly possible; for instance, choosing to minimize $V_0$ will instead produce the solutions with smallest value of the vacuum energy, which is perhaps what we want to find models matching the very small $V_0$ in our own Universe. There may be other better choices of objective function depending on the particular features of the problem one is dealing with; we only wish to emphasize that this standard feature of semidefinite programs can be used to find reliable solutions in the dS search.
\end{itemize}

We will now illustrate the ideas above in a simple toy model. We will not find a full de Sitter solution (otherwise, we would be writing the paper about that!); rather, we will take the simple example of a toroidal compactification, and fix by hand all moduli except for two---the volume $R$ and another modulus that we call $\phi$. Within this truncated problem, we will be able to apply the algorithm above to determine the regions in parameter space where a minimum in the $(R,\phi)$ directions may exist, producing exclusion plots similar to those now familiar in bootstrap. Finding an actual de Sitter would require doing the same taking into account all the moduli that we neglect; but finding no viable points in parameter space for $(R,\phi)$ would be enough to show that the scenario admits no dS minima.  

Our toy example is an M-theory compactification on $T^8$, down to $d=3$. Since $T^8$ has 36 metric moduli and M-theory does not have any modulus, the moduli we would need to stabilise to obtain a de Sitter minimum are $\{R,\phi^I\},\, I=1,...,35$. In line with the above, we will forget about 34 of these moduli and just analyse  a particular two-parameter submanifold of moduli space along which the metric only depends on $\{R,\phi\}$, 
\begin{equation}
    g = R^2\,\text{diag}(\phi^{1/4},\phi^{1/4},\phi^{1/4},\phi^{1/4},\phi^{1/4},\phi^{1/4},\phi^{1/4},\phi^{-7/4}) \,.
    \label{eq:metric}
\end{equation}

We can turn on $G_4$ and $G_7$ fluxes, so that the flux-Casimir potential in 3d takes the form 
\begin{equation}
    \frac{V_{\rm 3d}(R,\phi)}{(M_3)^3} \sim \frac{1}{R^{24}}\left(\frac{n_7^2 + \phi^2\,\tilde{n}_7^2}{R^{6}\,\phi^{7/4}} - \frac{\mathcal{C}(\phi)}{R^{3}} + \frac{n_4^2 + \phi^2\,\tilde{n}_4^2}{\phi}\right) \,,
\label{noname}\end{equation}
where $\tilde{n}_p$ has one leg along the $g_{88}$ direction, while $n_p$ has no leg along $g_{88}$.  These flux numbers are combinations of different fluxes with the same moduli dependence; for instance, there are $35$ four-cycles with no legs along $g_{88}$ and another $35$ with one leg along $g_{88}$. Due to the simple form of the metric $g$ \eq{eq:metric} there are only 4 distinct combinations of the moduli that arise in the flux potential and this is all we need to keep track of to run the semidefinite algorithm. However, one should keep in mind that the parameters $n_p,\,\tilde{n}_p$ do \emph{not} correspond to the properly quantised fluxes; as mentioned above, to declare that an actual de Sitter minimum exists, we would also need to take flux quantisation into account.    

We will also take advantage of a scaling symmetry in \eq{noname},
\begin{equation}
    \label{scal} 
    R\, \rightarrow \alpha\, R \,,
    \quad (n_7^2,\tilde{n}_7^2,n_4^2,\tilde{n}_4^2)\,\rightarrow\, (\alpha^3\, n_7^2,\alpha^3\,\tilde{n}_7^2,\alpha^{-3}\,n_4^2,\alpha^{-3}\,\tilde{n}_4^2) \,,
    \quad V_{\text{3d}}\,\rightarrow\, \alpha^{-27} \, V_{\text{3d}} \,,
\end{equation}
and use it to set $R=1$. The scaling symmetry \eq{scal} does not respect flux quantisation, but we are not worried about it at this stage. Once a solution is found for $R=1$, we can rescale back to obtain a dS for any value of $R$. Moreover, for the optimisation problem with a given $\phi = \phi_0$, we only need $\mathcal{C}(\phi_0)$ and the first two derivatives of the Casimir potential, 
\begin{equation}
    \mathcal{C}^{(1)} \equiv \phi\,\frac{\partial\log\Vcas}{\partial \phi}\bigg|_{\phi=\phi_0} \,,\quad\quad 
    \mathcal{C}^{(2)} \equiv \phi^2\,\frac{\partial^2\log\Vcas}{\partial \phi^2}\bigg|_{\phi=\phi_0} \,.
\end{equation}
We can then run the algorithm for different values of $(\mathcal{C}^{(1)},\mathcal{C}^{(2)})$ and obtain the minimum value of $\Vcas$  that allows for a de Sitter minimum. In this way, we have traded the moduli dependence by the two quantities $(\mathcal{C}^{(1)},\mathcal{C}^{(2)})$, which are the ones we try to ``bootstrap''.  In particular, there are regions of the $(\mathcal{C}^{(1)},\mathcal{C}^{(2)})$ plane that do not allow for a dS minimum at all---thus we can determine whether a dS minimum exists or not directly from properties of the Casimir energy $\mathcal{C}(\phi)$ alone. 

\begin{figure*}[t]
\centering
\includegraphics[height=5.7cm]{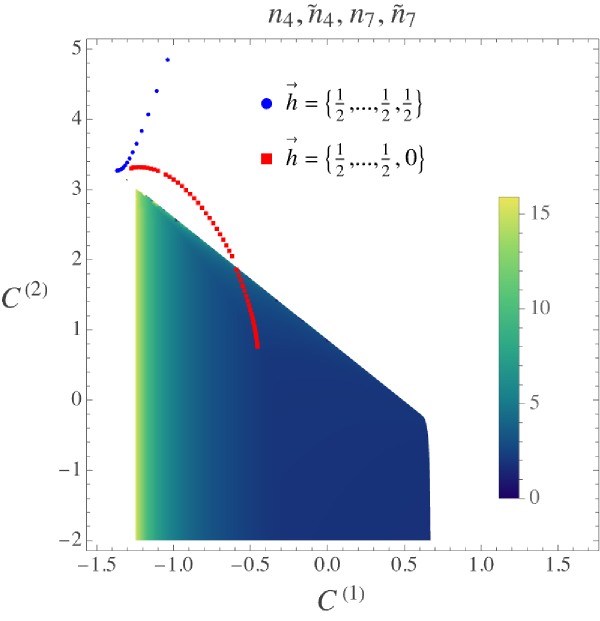}
\includegraphics[height=5.65cm]{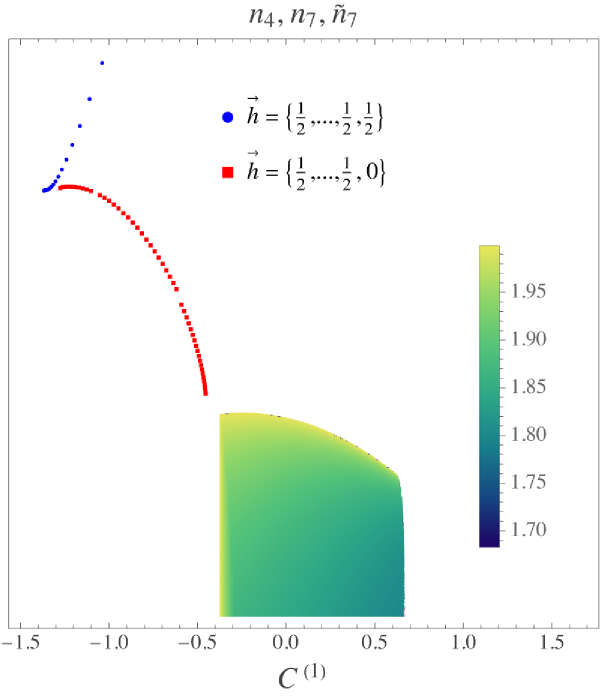}
\includegraphics[height=5.65cm]{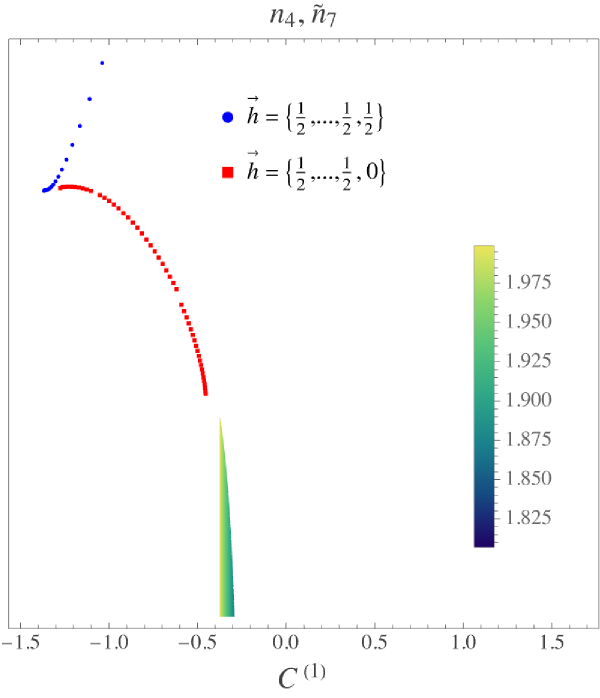}
\caption{Minimum value of $\Vcas$ that satisfies all constraints of our semidefinite optimisation problem \eq{diff} for each value of $(\mathcal{C}^{(1)},\mathcal{C}^{(2)})$, and different combinations of fluxes. White regions have no solution to the constraints and are therefore excluded. We also show the points corresponding to the Casimir potential $\mathcal{C}(\phi)$ computed for $\phi\in[0.5,2]$ and two different choices of spin structure on $T^8$, labeled using the notation of \cite{ValeixoBento:2025yhz}. The color in the allowed region represents the value of the objective function for the optimisation problem, the Casimir coefficient $\mathcal{C}$. Lower values/darker hues correspond to solutions which are under better control.}
\label{fig:sdo-nocasimir}
\end{figure*}

We scan the $(\mathcal{C}^{(1)},\mathcal{C}^{(2)})$ plane to find the region allowed by our ``bootstrap'' constraints---the region where dS minima may exist. This is completely analogous to e.g. the exclusion plots for conformal dimensions in the 3d Ising model in \cite{El-Showk:2012cjh}. We then check whether a given Casimir potential $\mathcal{C}(\phi)$, computed as described in \cite{ValeixoBento:2025yhz}, lands in the allowed region. This is analogous to plotting existing or putative CFT's in bootstrap plots. In \figref{fig:sdo-nocasimir}, we show these excluded regions for different sets of 4-form and 7-form fluxes, and with $\phi_0=2$. We can see that turning off fluxes reduces the allowed region of parameter space; for instance, switching off $\tilde{n}_4$ and $n_7$ simultaneously leaves a very narrow slit in parameter space where a dS minimum could exist. One the other hand, requiring specific types of fluxes to be turned on imposes conditions on the cycles of the RFM, which we can then use to help select an appropriate compactification manifold. We include in \figref{fig:sdo-nocasimir} the points corresponding to the Casimir potential $\mathcal{C}(\phi)$ computed for $\phi\in[0.5,2]$ and two different choices of spin structure on $T^8$, which we label using the notation of \cite{ValeixoBento:2025yhz}. While the fully twisted boundary conditions on $T^8$ lead to a Casimir potential that never lands on the allowed region, choosing periodic boundary conditions around the $g_{88}$ direction gives a potential that overlaps with this region for larger values of $\phi$, only when all fluxes are turned on. Although this appears to approach the allowed region in the remaining cases as well for larger values of $\phi$, we have checked that this does not happen. Finally, unlike traditional bootstrap plots, ours include a color grading that represents the value of the objective function $\mathcal{C}$; smaller values (darker hues) will lead, upon undoing the rescaling \eq{scal}, to solutions with larger $R$ and therefore better control.

\section{Summary and future directions} \label{s4}

In  \cite{ValeixoBento:2025yhz}, we showed that supersymmetry-breaking RFM's are a promising corner of the string Landscape, which can yield accelerating cosmologies and de Sitter critical points. The purpose of this note is to help organize the future exploration of this corner of the Landscape from a high-level point of view and, in particular, to help flesh out what is possible or not. To summarize our conclusions succintly, we have found that, within the landscape of RFM flux compactifications,
\begin{itemize}
    \item No dS$_4$ minimum exists,
    \item dS$_3$ minima \emph{may} exist, and
    \item the problem of finding dS$_3$ minima can be recast as the kind of \emph{semidefinite program} which is ubiquitous in the CFT bootstrap literature (see e.g. \cite{Simmons-Duffin:2015qma,Chester:2019wfx,Simmons-Duffin:2016gjk}).
\end{itemize}
To our knowledge, the last point has not been made in the literature so far (though the key idea of using linear programming techniques for flux vacua  comes from \cite{Grimm:2019ixq}, and the basic idea of linear inequalities for no-go theorems goes back to \cite{Flauger:2008ad}). The idea is simple but particularly fortunate, since the relevance of semidefinite programming for the bootstrap community has led to very rapid progress and an abundance of numerical tools which can now be repurposed for the search of dS minima (see \cite{Bonifacio:2020xoc,Bonifacio:2021msa} for other uses of these techniques beyond traditional CFT bootstrap). We also wish to emphasize that, although we centered the discussion on RFM compactifications, the features leading to a semidefinite program hold for more general flux compactifications e.g. on Calabi-Yau's, or including orientifolds. This is a question worth exploring further.

We end with a couple of observations. First, if a dS$_3$ minimum were to exist, it might not contradict the physical motivations of conjectures like \cite{Obied:2018sgi,Ooguri:2018wrx,Bedroya:2019snp}, since three-dimensional gravity is an edge case, where there are no propagating gravitons and  dynamical black hole formation does not take place. From this angle, the findings of the present paper somewhat align with the no dS conjectures, since we could very well have found a dS$_4$ in the RFM setup---now we have shown one cannot. Similar comments apply to dS$_2$, a setup which we have not analyzed in detail but expect to be even less constrained than dS$_3$. 

Our second point is that our no-go is strictly about RFM's. The lesson is that, if one is inclined to look for dS$_4$ minima along these lines, it might be necessary to generalize the setup beyond manifolds, e.g. to toroidal orbifolds---with all the complications that closed-string tachyons and Casimir energy calculations may entail.  Another avenue of generalization, suggested by the almost dS$_4$ we found in $D=9$, is to expand our search beyond strictly Riemann-flat vacua, e.g. to nilmanifolds or other homogeneous spaces where the Casimir energy may be computed explicitly, and which may admit metrics close enough to Ricci-flat so as to apply the perturbative techniques of \cite{ValeixoBento:2025yhz}. All in all, we remain interested in finding out whether there are RFM or RFM-like dS minima in String Theory, although for us this question now takes a backseat with respect to more general, time-dependent accelerating cosmologies (in light of the  recent DESI results).

\vspace{0.5cm}
\textbf{Acknowledgements:} We are indebted to 
Veronica Collazuol,  
Bernardo Fraiman,  
Andriana Makridou,  
Susha Parameswaran,  
Salvatore Raucci,  
Marco Serra and
Irene Valenzuela
for very valuable discussions and explanations.
MM thanks the SCGP for hospitality and a stimulating environment during the Summer '24 workshop, where parts of this workshop were completed. 
We also thank CERN and the Harvard Swampland
Initiative for hospitality during part of this work, and the
Aspen Center for Physics in the context of the 2025 Program on Holography, Bootstrap, and Swampland, which is supported by National Science Foundation grant PHY-2210452 and a grant from the Simons Foundation (1161654, Troyer), for providing a stimulating environment during
its completion.
The authors gratefully acknowledge the support of an Atraccion del Talento Fellowship 2022-T1/TIC-23956 from Comunidad de Madrid, which supported both authors in the early stages of this project, as well as the Spanish State Research Agency (Agencia Estatal de Investigacion) through the grants IFT Centro de Excelencia Severo Ochoa CEX2020-001007-S, PID2021-123017NB-I00, and Europa Excelencia EUR2024-153547. MM is currently supported by the RyC grant RYC2022-037545-I from AEI.

\bibliographystyle{utphys}
\bibliography{references}

\providecommand{\href}[2]{#2}\begingroup\raggedright\begin{thebibliography}{10}

\bibitem{SupernovaCosmologyProject:1998vns}
{\bf Supernova Cosmology Project} Collaboration, S.~Perlmutter {\em et al.}, ``{Measurements of $\Omega$ and $\Lambda$ from 42 High Redshift Supernovae}'', \href{http://dx.doi.org/10.1086/307221}{{\em Astrophys. J.} {\bf 517} (1999)  565--586}, \href{http://arxiv.org/abs/astro-ph/9812133}{{\tt arXiv:astro-ph/9812133}}.

\bibitem{SupernovaSearchTeam:1998fmf}
{\bf Supernova Search Team} Collaboration, A.~G. Riess {\em et al.}, ``{Observational evidence from supernovae for an accelerating universe and a cosmological constant}'', \href{http://dx.doi.org/10.1086/300499}{{\em Astron. J.} {\bf 116} (1998)  1009--1038}, \href{http://arxiv.org/abs/astro-ph/9805201}{{\tt arXiv:astro-ph/9805201}}.

\bibitem{Dine:1985he}
M.~Dine and N.~Seiberg, ``{Is the Superstring Weakly Coupled?}'', \href{http://dx.doi.org/10.1016/0370-2693(85)90927-X}{{\em Phys. Lett. B} {\bf 162} (1985)  299--302}.

\bibitem{Ooguri:2006in}
H.~Ooguri and C.~Vafa, ``{On the Geometry of the String Landscape and the Swampland}'', \href{http://dx.doi.org/10.1016/j.nuclphysb.2006.10.033}{{\em Nucl. Phys. B} {\bf 766} (2007)  21--33}, \href{http://arxiv.org/abs/hep-th/0605264}{{\tt arXiv:hep-th/0605264}}.

\bibitem{Lee:2019wij}
S.-J. Lee, W.~Lerche, and T.~Weigand, ``{Emergent strings from infinite distance limits}'', \href{http://dx.doi.org/10.1007/JHEP02(2022)190}{{\em JHEP} {\bf 02} (2022)  190}, \href{http://arxiv.org/abs/1910.01135}{{\tt arXiv:1910.01135 [hep-th]}}.

\bibitem{Palti:2019pca}
E.~Palti, ``{The Swampland: Introduction and Review}'', \href{http://dx.doi.org/10.1002/prop.201900037}{{\em Fortsch. Phys.} {\bf 67} (2019) no.~6, 1900037}, \href{http://arxiv.org/abs/1903.06239}{{\tt arXiv:1903.06239 [hep-th]}}.

\bibitem{vanBeest:2021lhn}
M.~van Beest, J.~Calder\'on-Infante, D.~Mirfendereski, and I.~Valenzuela, ``{Lectures on the Swampland Program in String Compactifications}'', \href{http://dx.doi.org/10.1016/j.physrep.2022.09.002}{{\em Phys. Rept.} {\bf 989} (2022)  1--50}, \href{http://arxiv.org/abs/2102.01111}{{\tt arXiv:2102.01111 [hep-th]}}.

\bibitem{Danielsson:2018ztv}
U.~H. Danielsson and T.~Van~Riet, ``{What if string theory has no de Sitter vacua?}'', \href{http://dx.doi.org/10.1142/S0218271818300070}{{\em Int. J. Mod. Phys. D} {\bf 27} (2018) no.~12, 1830007}, \href{http://arxiv.org/abs/1804.01120}{{\tt arXiv:1804.01120 [hep-th]}}.

\bibitem{Obied:2018sgi}
G.~Obied, H.~Ooguri, L.~Spodyneiko, and C.~Vafa, ``{De Sitter Space and the Swampland}'', \href{http://arxiv.org/abs/1806.08362}{{\tt arXiv:1806.08362 [hep-th]}}.

\bibitem{Ooguri:2018wrx}
H.~Ooguri, E.~Palti, G.~Shiu, and C.~Vafa, ``{Distance and de Sitter Conjectures on the Swampland}'', \href{http://dx.doi.org/10.1016/j.physletb.2018.11.018}{{\em Phys. Lett. B} {\bf 788} (2019)  180--184}, \href{http://arxiv.org/abs/1810.05506}{{\tt arXiv:1810.05506 [hep-th]}}.

\bibitem{Andriot:2018wzk}
D.~Andriot, ``{On the de Sitter swampland criterion}'', \href{http://dx.doi.org/10.1016/j.physletb.2018.09.022}{{\em Phys. Lett. B} {\bf 785} (2018)  570--573}, \href{http://arxiv.org/abs/1806.10999}{{\tt arXiv:1806.10999 [hep-th]}}.

\bibitem{Andriot:2018mav}
D.~Andriot and C.~Roupec, ``{Further refining the de Sitter swampland conjecture}'', \href{http://dx.doi.org/10.1002/prop.201800105}{{\em Fortsch. Phys.} {\bf 67} (2019) no.~1-2, 1800105}, \href{http://arxiv.org/abs/1811.08889}{{\tt arXiv:1811.08889 [hep-th]}}.

\bibitem{Bedroya:2019snp}
A.~Bedroya and C.~Vafa, ``{Trans-Planckian Censorship and the Swampland}'', \href{http://dx.doi.org/10.1007/JHEP09(2020)123}{{\em JHEP} {\bf 09} (2020)  123}, \href{http://arxiv.org/abs/1909.11063}{{\tt arXiv:1909.11063 [hep-th]}}.

\bibitem{Grimm:2019ixq}
T.~W. Grimm, C.~Li, and I.~Valenzuela, ``{Asymptotic Flux Compactifications and the Swampland}'', \href{http://dx.doi.org/10.1007/JHEP06(2020)009}{{\em JHEP} {\bf 06} (2020)  009}, \href{http://arxiv.org/abs/1910.09549}{{\tt arXiv:1910.09549 [hep-th]}}. [Erratum: JHEP 01, 007 (2021)].

\bibitem{Calderon-Infante:2022nxb}
J.~Calder\'on-Infante, I.~Ruiz, and I.~Valenzuela, ``{Asymptotic accelerated expansion in string theory and the Swampland}'', \href{http://dx.doi.org/10.1007/JHEP06(2023)129}{{\em JHEP} {\bf 06} (2023)  129}, \href{http://arxiv.org/abs/2209.11821}{{\tt arXiv:2209.11821 [hep-th]}}.

\bibitem{Etheredge:2024tok}
M.~Etheredge, B.~Heidenreich, T.~Rudelius, I.~Ruiz, and I.~Valenzuela, ``{Taxonomy of infinite distance limits}'', \href{http://dx.doi.org/10.1007/JHEP03(2025)213}{{\em JHEP} {\bf 03} (2025)  213}, \href{http://arxiv.org/abs/2405.20332}{{\tt arXiv:2405.20332 [hep-th]}}.

\bibitem{Kachru:2003aw}
S.~Kachru, R.~Kallosh, A.~D. Linde, and S.~P. Trivedi, ``{De Sitter vacua in string theory}'', \href{http://dx.doi.org/10.1103/PhysRevD.68.046005}{{\em Phys. Rev. D} {\bf 68} (2003)  046005}, \href{http://arxiv.org/abs/hep-th/0301240}{{\tt arXiv:hep-th/0301240}}.

\bibitem{Balasubramanian:2005zx}
V.~Balasubramanian, P.~Berglund, J.~P. Conlon, and F.~Quevedo, ``{Systematics of moduli stabilisation in Calabi-Yau flux compactifications}'', \href{http://dx.doi.org/10.1088/1126-6708/2005/03/007}{{\em JHEP} {\bf 03} (2005)  007}, \href{http://arxiv.org/abs/hep-th/0502058}{{\tt arXiv:hep-th/0502058}}.

\bibitem{Gao:2020xqh}
X.~Gao, A.~Hebecker, and D.~Junghans, ``{Control issues of KKLT}'', \href{http://dx.doi.org/10.1002/prop.202000089}{{\em Fortsch. Phys.} {\bf 68} (2020)  2000089}, \href{http://arxiv.org/abs/2009.03914}{{\tt arXiv:2009.03914 [hep-th]}}.

\bibitem{Demirtas:2021nlu}
M.~Demirtas, M.~Kim, L.~McAllister, J.~Moritz, and A.~Rios-Tascon, ``{Small cosmological constants in string theory}'', \href{http://dx.doi.org/10.1007/JHEP12(2021)136}{{\em JHEP} {\bf 12} (2021)  136}, \href{http://arxiv.org/abs/2107.09064}{{\tt arXiv:2107.09064 [hep-th]}}.

\bibitem{Junghans:2022exo}
D.~Junghans, ``{LVS de Sitter vacua are probably in the swampland}'', \href{http://dx.doi.org/10.1016/j.nuclphysb.2023.116179}{{\em Nucl. Phys. B} {\bf 990} (2023)  116179}, \href{http://arxiv.org/abs/2201.03572}{{\tt arXiv:2201.03572 [hep-th]}}.

\bibitem{Bena:2022cwb}
I.~Bena, E.~Dudas, M.~Gra\~na, G.~Lo~Monaco, and D.~Toulikas, ``{Bare-bones de Sitter vacua}'', \href{http://dx.doi.org/10.1103/PhysRevD.108.L021901}{{\em Phys. Rev. D} {\bf 108} (2023) no.~2, L021901}, \href{http://arxiv.org/abs/2202.02327}{{\tt arXiv:2202.02327 [hep-th]}}.

\bibitem{Gao:2022fdi}
X.~Gao, A.~Hebecker, S.~Schreyer, and V.~Venken, ``{The LVS parametric tadpole constraint}'', \href{http://dx.doi.org/10.1007/JHEP07(2022)056}{{\em JHEP} {\bf 07} (2022)  056}, \href{http://arxiv.org/abs/2202.04087}{{\tt arXiv:2202.04087 [hep-th]}}.

\bibitem{Lust:2022lfc}
S.~L\"ust, C.~Vafa, M.~Wiesner, and K.~Xu, ``{Holography and the KKLT scenario}'', \href{http://dx.doi.org/10.1007/JHEP10(2022)188}{{\em JHEP} {\bf 10} (2022)  188}, \href{http://arxiv.org/abs/2204.07171}{{\tt arXiv:2204.07171 [hep-th]}}.

\bibitem{Lust:2022xoq}
S.~L\"ust and L.~Randall, ``{Effective Theory of Warped Compactifications and the Implications for KKLT}'', \href{http://dx.doi.org/10.1002/prop.202200103}{{\em Fortsch. Phys.} {\bf 70} (2022) no.~7-8, 2200103}, \href{http://arxiv.org/abs/2206.04708}{{\tt arXiv:2206.04708 [hep-th]}}.

\bibitem{Hebecker:2022zme}
A.~Hebecker, S.~Schreyer, and V.~Venken, ``{Curvature corrections to KPV: do we need deep throats?}'', \href{http://dx.doi.org/10.1007/JHEP10(2022)166}{{\em JHEP} {\bf 10} (2022)  166}, \href{http://arxiv.org/abs/2208.02826}{{\tt arXiv:2208.02826 [hep-th]}}.

\bibitem{Bena:2022ive}
I.~Bena, E.~Dudas, M.~Gra\~na, G.~Lo~Monaco, and D.~Toulikas, ``{$ \overline{\textrm{D}3} $-branes and gaugino condensation}'', \href{http://dx.doi.org/10.1007/JHEP12(2023)019}{{\em JHEP} {\bf 12} (2023)  019}, \href{http://arxiv.org/abs/2211.14381}{{\tt arXiv:2211.14381 [hep-th]}}.

\bibitem{ValeixoBento:2023nbv}
B.~Valeixo~Bento, D.~Chakraborty, S.~Parameswaran, and I.~Zavala, ``{De Sitter vacua \textemdash{} when are \textquoteleft{}subleading corrections\textquoteright{} really subleading?}'', \href{http://dx.doi.org/10.1007/JHEP11(2023)075}{{\em JHEP} {\bf 11} (2023)  075}, \href{http://arxiv.org/abs/2306.07332}{{\tt arXiv:2306.07332 [hep-th]}}.

\bibitem{McAllister:2024lnt}
L.~McAllister, J.~Moritz, R.~Nally, and A.~Schachner, ``{Candidate de Sitter vacua}'', \href{http://dx.doi.org/10.1103/PhysRevD.111.086015}{{\em Phys. Rev. D} {\bf 111} (2025) no.~8, 086015}, \href{http://arxiv.org/abs/2406.13751}{{\tt arXiv:2406.13751 [hep-th]}}.

\bibitem{Kim:2024dnw}
M.~Kim, ``{String perturbation theory of Klebanov-Strassler throat}'', \href{http://arxiv.org/abs/2409.19048}{{\tt arXiv:2409.19048 [hep-th]}}.

\bibitem{Moritz:2025bsi}
J.~Moritz, ``{$G_2$-manifolds from Diophantine equations}'', \href{http://arxiv.org/abs/2505.15883}{{\tt arXiv:2505.15883 [hep-th]}}.

\bibitem{Gibbons:1984kp}
G.~W. Gibbons, ``{ASPECTS OF SUPERGRAVITY THEORIES}'', in {\em {XV GIFT Seminar on Supersymmetry and Supergravity}}.
\newblock 6, 1984.

\bibitem{Maldacena:2000mw}
J.~M. Maldacena and C.~Nunez, ``{Supergravity description of field theories on curved manifolds and a no go theorem}'', \href{http://dx.doi.org/10.1142/S0217751X01003937}{{\em Int. J. Mod. Phys. A} {\bf 16} (2001)  822--855}, \href{http://arxiv.org/abs/hep-th/0007018}{{\tt arXiv:hep-th/0007018}}.

\bibitem{ValeixoBento:2025yhz}
B.~Valeixo~Bento and M.~Montero, ``{An M-theory dS maximum from Casimir energies on Riemann-flat manifolds}'', \href{http://arxiv.org/abs/2507.02037}{{\tt arXiv:2507.02037 [hep-th]}}.

\bibitem{DeLuca:2021pej}
G.~B. De~Luca, E.~Silverstein, and G.~Torroba, ``{Hyperbolic compactification of M-theory and de Sitter quantum gravity}'', \href{http://dx.doi.org/10.21468/SciPostPhys.12.3.083}{{\em SciPost Phys.} {\bf 12} (2022) no.~3, 083}, \href{http://arxiv.org/abs/2104.13380}{{\tt arXiv:2104.13380 [hep-th]}}.

\bibitem{Vafa:1991uz}
C.~Vafa, ``{Topological mirrors and quantum rings}'', {\em AMS/IP Stud. Adv. Math.} {\bf 9} (1998)  97--120, \href{http://arxiv.org/abs/hep-th/9111017}{{\tt arXiv:hep-th/9111017}}.

\bibitem{Candelas:1993nd}
P.~Candelas, E.~Derrick, and L.~Parkes, ``{Generalized Calabi-Yau manifolds and the mirror of a rigid manifold}'', \href{http://dx.doi.org/10.1016/0550-3213(93)90276-U}{{\em Nucl. Phys. B} {\bf 407} (1993)  115--154}, \href{http://arxiv.org/abs/hep-th/9304045}{{\tt arXiv:hep-th/9304045}}.

\bibitem{Vafa:1989xc}
C.~Vafa, ``{String Vacua and Orbifoldized L-G Models}'', \href{http://dx.doi.org/10.1142/S0217732389001350}{{\em Mod. Phys. Lett. A} {\bf 4} (1989)  1169}.

\bibitem{Becker:2006ks}
K.~Becker, M.~Becker, C.~Vafa, and J.~Walcher, ``{Moduli Stabilization in Non-Geometric Backgrounds}'', \href{http://dx.doi.org/10.1016/j.nuclphysb.2007.01.034}{{\em Nucl. Phys. B} {\bf 770} (2007)  1--46}, \href{http://arxiv.org/abs/hep-th/0611001}{{\tt arXiv:hep-th/0611001}}.

\bibitem{Bardzell:2022jfh}
J.~Bardzell, E.~Gonzalo, M.~Rajaguru, D.~Smith, and T.~Wrase, ``{Type IIB flux compactifications with h$^{1,1}$ = 0}'', \href{http://dx.doi.org/10.1007/JHEP06(2022)166}{{\em JHEP} {\bf 06} (2022)  166}, \href{http://arxiv.org/abs/2203.15818}{{\tt arXiv:2203.15818 [hep-th]}}.

\bibitem{Becker:2022hse}
K.~Becker, E.~Gonzalo, J.~Walcher, and T.~Wrase, ``{Fluxes, vacua, and tadpoles meet Landau-Ginzburg and Fermat}'', \href{http://dx.doi.org/10.1007/JHEP12(2022)083}{{\em JHEP} {\bf 12} (2022)  083}, \href{http://arxiv.org/abs/2210.03706}{{\tt arXiv:2210.03706 [hep-th]}}.

\bibitem{Cremonini:2023suw}
S.~Cremonini, E.~Gonzalo, M.~Rajaguru, Y.~Tang, and T.~Wrase, ``{On asymptotic dark energy in string theory}'', \href{http://dx.doi.org/10.1007/JHEP09(2023)075}{{\em JHEP} {\bf 09} (2023)  075}, \href{http://arxiv.org/abs/2306.15714}{{\tt arXiv:2306.15714 [hep-th]}}.

\bibitem{Becker:2024ijy}
K.~Becker, M.~Rajaguru, A.~Sengupta, J.~Walcher, and T.~Wrase, ``{Stabilizing massless fields with fluxes in Landau-Ginzburg models}'', \href{http://dx.doi.org/10.1007/JHEP08(2024)069}{{\em JHEP} {\bf 08} (2024)  069}, \href{http://arxiv.org/abs/2406.03435}{{\tt arXiv:2406.03435 [hep-th]}}.

\bibitem{Chen:2025rkb}
S.~Chen, D.~van~de Heisteeg, and C.~Vafa, ``{Symmetries and M-theory-like Vacua in Four Dimensions}'', \href{http://arxiv.org/abs/2503.16599}{{\tt arXiv:2503.16599 [hep-th]}}.

\bibitem{Parameswaran:2024mrc}
S.~Parameswaran and M.~Serra, ``{On (A)dS solutions from Scherk-Schwarz orbifolds}'', \href{http://dx.doi.org/10.1007/JHEP10(2024)039}{{\em JHEP} {\bf 10} (2024)  039}, \href{http://arxiv.org/abs/2407.16781}{{\tt arXiv:2407.16781 [hep-th]}}.

\bibitem{Mertens:2022irh}
T.~G. Mertens and G.~J. Turiaci, ``{Solvable models of quantum black holes: a review on Jackiw{\textendash}Teitelboim gravity}'', \href{http://dx.doi.org/10.1007/s41114-023-00046-1}{{\em Living Rev. Rel.} {\bf 26} (2023) no.~1, 4}, \href{http://arxiv.org/abs/2210.10846}{{\tt arXiv:2210.10846 [hep-th]}}.

\bibitem{Maldacena:2019cbz}
J.~Maldacena, G.~J. Turiaci, and Z.~Yang, ``{Two dimensional Nearly de Sitter gravity}'', \href{http://dx.doi.org/10.1007/JHEP01(2021)139}{{\em JHEP} {\bf 01} (2021)  139}, \href{http://arxiv.org/abs/1904.01911}{{\tt arXiv:1904.01911 [hep-th]}}.

\bibitem{Iliesiu:2020zld}
L.~V. Iliesiu, J.~Kruthoff, G.~J. Turiaci, and H.~Verlinde, ``{JT gravity at finite cutoff}'', \href{http://dx.doi.org/10.21468/SciPostPhys.9.2.023}{{\em SciPost Phys.} {\bf 9} (2020)  023}, \href{http://arxiv.org/abs/2004.07242}{{\tt arXiv:2004.07242 [hep-th]}}.

\bibitem{Nanda:2023wne}
K.~K. Nanda, S.~K. Sake, and S.~P. Trivedi, ``{JT gravity in de Sitter space and the problem of time}'', \href{http://dx.doi.org/10.1007/JHEP02(2024)145}{{\em JHEP} {\bf 02} (2024)  145}, \href{http://arxiv.org/abs/2307.15900}{{\tt arXiv:2307.15900 [hep-th]}}.

\bibitem{Weinberg:2000cr}
S.~Weinberg, {\em {The quantum theory of fields. Vol. 3: Supersymmetry}}.
\newblock Cambridge University Press, 6, 2013.

\bibitem{Freedman:2012zz}
D.~Z. Freedman and A.~Van~Proeyen, \href{http://dx.doi.org/10.1017/CBO9781139026833}{{\em {Supergravity}}}.
\newblock Cambridge Univ. Press, Cambridge, UK, 5, 2012.

\bibitem{Heidenreich:2015nta}
B.~Heidenreich, M.~Reece, and T.~Rudelius, ``{Sharpening the Weak Gravity Conjecture with Dimensional Reduction}'', \href{http://dx.doi.org/10.1007/JHEP02(2016)140}{{\em JHEP} {\bf 02} (2016)  140}, \href{http://arxiv.org/abs/1509.06374}{{\tt arXiv:1509.06374 [hep-th]}}.

\bibitem{Polchinskiv2}
J.~Polchinski, \href{http://dx.doi.org/10.1017/CBO9780511618123}{{\em {String theory. Vol. 2: Superstring theory and beyond}}}.
\newblock Cambridge Monographs on Mathematical Physics. Cambridge University Press, 12, 2007.

\bibitem{Scherk:1978ta}
J.~Scherk and J.~H. Schwarz, ``{Spontaneous Breaking of Supersymmetry Through Dimensional Reduction}'', \href{http://dx.doi.org/10.1016/0370-2693(79)90425-8}{{\em Phys. Lett. B} {\bf 82} (1979)  60--64}.

\bibitem{Aharony:2016kai}
O.~Aharony, Y.~Tachikawa, and K.~Gomi, ``{S-folds and 4d N=3 superconformal field theories}'', \href{http://dx.doi.org/10.1007/JHEP06(2016)044}{{\em JHEP} {\bf 06} (2016)  044}, \href{http://arxiv.org/abs/1602.08638}{{\tt arXiv:1602.08638 [hep-th]}}.

\bibitem{Etheredge:2023ler}
M.~Etheredge, I.~Garcia~Etxebarria, B.~Heidenreich, and S.~Rauch, ``{Branes and symmetries for $ \mathcal{N} $ = 3 S-folds}'', \href{http://dx.doi.org/10.1007/JHEP09(2023)005}{{\em JHEP} {\bf 09} (2023)  005}, \href{http://arxiv.org/abs/2302.14068}{{\tt arXiv:2302.14068 [hep-th]}}.

\bibitem{Narain:1985jj}
K.~S. Narain, ``{New Heterotic String Theories in Uncompactified Dimensions {\ensuremath{<}} 10}'', \href{http://dx.doi.org/10.1016/0370-2693(86)90682-9}{{\em Phys. Lett. B} {\bf 169} (1986)  41--46}.

\bibitem{Narain:1986am}
K.~S. Narain, M.~H. Sarmadi, and E.~Witten, ``{A Note on Toroidal Compactification of Heterotic String Theory}'', \href{http://dx.doi.org/10.1016/0550-3213(87)90001-0}{{\em Nucl. Phys. B} {\bf 279} (1987)  369--379}.

\bibitem{Giveon:1988tt}
A.~Giveon, E.~Rabinovici, and G.~Veneziano, ``{Duality in String Background Space}'', \href{http://dx.doi.org/10.1016/0550-3213(89)90489-6}{{\em Nucl. Phys. B} {\bf 322} (1989)  167--184}.

\bibitem{Giveon:1994fu}
A.~Giveon, M.~Porrati, and E.~Rabinovici, ``{Target space duality in string theory}'', \href{http://dx.doi.org/10.1016/0370-1573(94)90070-1}{{\em Phys. Rept.} {\bf 244} (1994)  77--202}, \href{http://arxiv.org/abs/hep-th/9401139}{{\tt arXiv:hep-th/9401139}}.

\bibitem{Fraiman:2018ebo}
B.~Fraiman, M.~Gra{\~n}a, and C.~A. N{\'u}{\~n}ez, ``{A new twist on heterotic string compactifications}'', \href{http://dx.doi.org/10.1007/JHEP09(2018)078}{{\em JHEP} {\bf 09} (2018)  078}, \href{http://arxiv.org/abs/1805.11128}{{\tt arXiv:1805.11128 [hep-th]}}.

\bibitem{Aharony:2007du}
O.~Aharony, Z.~Komargodski, and A.~Patir, ``{The Moduli space and M(atrix) theory of 9d N=1 backgrounds of M/string theory}'', \href{http://dx.doi.org/10.1088/1126-6708/2007/05/073}{{\em JHEP} {\bf 05} (2007)  073}, \href{http://arxiv.org/abs/hep-th/0702195}{{\tt arXiv:hep-th/0702195}}.

\bibitem{lambert2025closed}
T.~P. Lambert, J.~G. Ratcliffe, and S.~T. Tschantz, ``Closed flat riemannian 4-manifolds'', {\em Annales math{\'e}matiques du Qu{\'e}bec} (2025)  1--47.

\bibitem{ValeixoBento:2025emh}
B.~Valeixo~Bento, D.~Chakraborty, S.~Parameswaran, and I.~Zavala, ``{A guide to frames, 2{\ensuremath{\pi}}{\textquoteright}s, scales and corrections in string compactifications}'', \href{http://dx.doi.org/10.1142/S0218271825300034}{{\em Int. J. Mod. Phys. D} {\bf 34} (2025) no.~10, 2530003}, \href{http://arxiv.org/abs/2301.05178}{{\tt arXiv:2301.05178 [hep-th]}}.

\bibitem{Bergshoeff:2002nv}
E.~Bergshoeff, T.~de~Wit, U.~Gran, R.~Linares, and D.~Roest, ``{(Non)Abelian gauged supergravities in nine-dimensions}'', \href{http://dx.doi.org/10.1088/1126-6708/2002/10/061}{{\em JHEP} {\bf 10} (2002)  061}, \href{http://arxiv.org/abs/hep-th/0209205}{{\tt arXiv:hep-th/0209205}}.

\bibitem{Fernandez-Melgarejo:2011nso}
J.~J. Fernandez-Melgarejo, T.~Ortin, and E.~Torrente-Lujan, ``{The general gaugings of maximal d=9 supergravity}'', \href{http://dx.doi.org/10.1007/JHEP10(2011)068}{{\em JHEP} {\bf 10} (2011)  068}, \href{http://arxiv.org/abs/1106.1760}{{\tt arXiv:1106.1760 [hep-th]}}.

\bibitem{Bergshoeff:1995as}
E.~Bergshoeff, C.~M. Hull, and T.~Ortin, ``{Duality in the type II superstring effective action}'', \href{http://dx.doi.org/10.1016/0550-3213(95)00367-2}{{\em Nucl. Phys. B} {\bf 451} (1995)  547--578}, \href{http://arxiv.org/abs/hep-th/9504081}{{\tt arXiv:hep-th/9504081}}.

\bibitem{shin1998group}
J.~Shin, ``Group actions on the 3-dimensional nilmanifold'', {\em Trends in Mathematics} {\bf 1} (1998)  62.

\bibitem{Kaloper:1999yr}
N.~Kaloper and R.~C. Myers, ``{The Odd story of massive supergravity}'', \href{http://dx.doi.org/10.1088/1126-6708/1999/05/010}{{\em JHEP} {\bf 05} (1999)  010}, \href{http://arxiv.org/abs/hep-th/9901045}{{\tt arXiv:hep-th/9901045}}.

\bibitem{Kachru:2002sk}
S.~Kachru, M.~B. Schulz, P.~K. Tripathy, and S.~P. Trivedi, ``{New supersymmetric string compactifications}'', \href{http://dx.doi.org/10.1088/1126-6708/2003/03/061}{{\em JHEP} {\bf 03} (2003)  061}, \href{http://arxiv.org/abs/hep-th/0211182}{{\tt arXiv:hep-th/0211182}}.

\bibitem{Grana:2006kf}
M.~Grana, R.~Minasian, M.~Petrini, and A.~Tomasiello, ``{A Scan for new N=1 vacua on twisted tori}'', \href{http://dx.doi.org/10.1088/1126-6708/2007/05/031}{{\em JHEP} {\bf 05} (2007)  031}, \href{http://arxiv.org/abs/hep-th/0609124}{{\tt arXiv:hep-th/0609124}}.

\bibitem{Andriot:2015sia}
D.~Andriot, ``{New supersymmetric vacua on solvmanifolds}'', \href{http://dx.doi.org/10.1007/JHEP02(2016)112}{{\em JHEP} {\bf 02} (2016)  112}, \href{http://arxiv.org/abs/1507.00014}{{\tt arXiv:1507.00014 [hep-th]}}.

\bibitem{Andriot:2018tmb}
D.~Andriot and D.~Tsimpis, ``{Laplacian spectrum on a nilmanifold, truncations and effective theories}'', \href{http://dx.doi.org/10.1007/JHEP09(2018)096}{{\em JHEP} {\bf 09} (2018)  096}, \href{http://arxiv.org/abs/1806.05156}{{\tt arXiv:1806.05156 [hep-th]}}.

\bibitem{belegradek2002metrics}
I.~Belegradek and G.~Wei, ``Metrics of positive ricci curvature on vector bundles over nilmanifolds'', {\em Geometric \& Functional Analysis GAFA} {\bf 12} (2002) no.~1, 56--72.

\bibitem{belegradek2020iterated}
I.~Belegradek, ``Iterated circle bundles and infranilmanifolds'',.

\bibitem{Dvali:2005an}
G.~Dvali, ``{Three-form gauging of axion symmetries and gravity}'', \href{http://arxiv.org/abs/hep-th/0507215}{{\tt arXiv:hep-th/0507215}}.

\bibitem{Kaloper:2008fb}
N.~Kaloper and L.~Sorbo, ``{A Natural Framework for Chaotic Inflation}'', \href{http://dx.doi.org/10.1103/PhysRevLett.102.121301}{{\em Phys. Rev. Lett.} {\bf 102} (2009)  121301}, \href{http://arxiv.org/abs/0811.1989}{{\tt arXiv:0811.1989 [hep-th]}}.

\bibitem{Kaloper:2008qs}
N.~Kaloper and L.~Sorbo, ``{Where in the String Landscape is Quintessence}'', \href{http://dx.doi.org/10.1103/PhysRevD.79.043528}{{\em Phys. Rev. D} {\bf 79} (2009)  043528}, \href{http://arxiv.org/abs/0810.5346}{{\tt arXiv:0810.5346 [hep-th]}}.

\bibitem{Alonso-Alberca:2000wkg}
N.~Alonso-Alberca, P.~Meessen, and T.~Ortin, ``{An Sl(3,Z) multiplet of eight-dimensional type II supergravity theories and the gauged supergravity inside}'', \href{http://dx.doi.org/10.1016/S0550-3213(01)00110-9}{{\em Nucl. Phys. B} {\bf 602} (2001)  329--345}, \href{http://arxiv.org/abs/hep-th/0012032}{{\tt arXiv:hep-th/0012032}}.

\bibitem{LassoAndino:2016lwl}
{\'O}.~Lasso~Andino and T.~Ort{\'\i}n, ``{On gauged maximal $d = 8$ supergravities}'', \href{http://dx.doi.org/10.1088/1361-6382/aaafa9}{{\em Class. Quant. Grav.} {\bf 35} (2018) no.~7, 075011}, \href{http://arxiv.org/abs/1605.09629}{{\tt arXiv:1605.09629 [hep-th]}}.

\bibitem{Sezgin:2023hkc}
E.~Sezgin, ``{Survey of supergravities}'', \href{http://arxiv.org/abs/2312.06754}{{\tt arXiv:2312.06754 [hep-th]}}.

\bibitem{Cremmer:1997ct}
E.~Cremmer, B.~Julia, H.~Lu, and C.~N. Pope, ``{Dualization of dualities. 1.}'', \href{http://dx.doi.org/10.1016/S0550-3213(98)00136-9}{{\em Nucl. Phys. B} {\bf 523} (1998)  73--144}, \href{http://arxiv.org/abs/hep-th/9710119}{{\tt arXiv:hep-th/9710119}}.

\bibitem{Sezgin:1982gi}
E.~Sezgin and A.~Salam, ``{Maximal Extended Supergravity Theory in Seven-dimensions}'', \href{http://dx.doi.org/10.1016/0370-2693(82)90204-0}{{\em Phys. Lett. B} {\bf 118} (1982)  359}.

\bibitem{Obers:1998fb}
N.~A. Obers and B.~Pioline, ``{U duality and M theory}'', \href{http://dx.doi.org/10.1016/S0370-1573(99)00004-6}{{\em Phys. Rept.} {\bf 318} (1999)  113--225}, \href{http://arxiv.org/abs/hep-th/9809039}{{\tt arXiv:hep-th/9809039}}.

\bibitem{Samtleben:2008pe}
H.~Samtleben, ``{Lectures on Gauged Supergravity and Flux Compactifications}'', \href{http://dx.doi.org/10.1088/0264-9381/25/21/214002}{{\em Class. Quant. Grav.} {\bf 25} (2008)  214002}, \href{http://arxiv.org/abs/0808.4076}{{\tt arXiv:0808.4076 [hep-th]}}.

\bibitem{Braeger:2025kra}
N.~Braeger, A.~Debray, M.~Dierigl, J.~J. Heckman, and M.~Montero, ``{Cobordism Utopia: U-Dualities, Bordisms, and the Swampland}'', \href{http://arxiv.org/abs/2505.15885}{{\tt arXiv:2505.15885 [hep-th]}}.

\bibitem{Simmons-Duffin:2015qma}
D.~Simmons-Duffin, ``{A Semidefinite Program Solver for the Conformal Bootstrap}'', \href{http://dx.doi.org/10.1007/JHEP06(2015)174}{{\em JHEP} {\bf 06} (2015)  174}, \href{http://arxiv.org/abs/1502.02033}{{\tt arXiv:1502.02033 [hep-th]}}.

\bibitem{Chester:2019wfx}
S.~M. Chester, ``{Weizmann lectures on the numerical conformal bootstrap}'', \href{http://dx.doi.org/10.1016/j.physrep.2023.10.008}{{\em Phys. Rept.} {\bf 1045} (2023)  1--44}, \href{http://arxiv.org/abs/1907.05147}{{\tt arXiv:1907.05147 [hep-th]}}.

\bibitem{Simmons-Duffin:2016gjk}
D.~Simmons-Duffin, \href{http://dx.doi.org/10.1142/9789813149441_0001}{``{The Conformal Bootstrap}'',} in {\em {Theoretical Advanced Study Institute in Elementary Particle Physics}: {New Frontiers in Fields and Strings}}, pp.~1--74.
\newblock 2017.
\newblock \href{http://arxiv.org/abs/1602.07982}{{\tt arXiv:1602.07982 [hep-th]}}.

\bibitem{DallAgata:2025jii}
G.~Dall'Agata and F.~Zwirner, ``{Supersymmetry-breaking compactifications on Riemann-flat manifolds}'', \href{http://arxiv.org/abs/2507.02339}{{\tt arXiv:2507.02339 [hep-th]}}.

\bibitem{stiemke1915positive}
E.~Stiemke, ``{\"U}ber positive l{\"o}sungen homogener linearer gleichungen'', {\em Mathematische Annalen} {\bf 76} (1915) no.~2, 340--342.

\bibitem{El-Showk:2012cjh}
S.~El-Showk, M.~F. Paulos, D.~Poland, S.~Rychkov, D.~Simmons-Duffin, and A.~Vichi, ``{Solving the 3D Ising Model with the Conformal Bootstrap}'', \href{http://dx.doi.org/10.1103/PhysRevD.86.025022}{{\em Phys. Rev. D} {\bf 86} (2012)  025022}, \href{http://arxiv.org/abs/1203.6064}{{\tt arXiv:1203.6064 [hep-th]}}.

\bibitem{Flauger:2008ad}
R.~Flauger, S.~Paban, D.~Robbins, and T.~Wrase, ``{Searching for slow-roll moduli inflation in massive type IIA supergravity with metric fluxes}'', \href{http://dx.doi.org/10.1103/PhysRevD.79.086011}{{\em Phys. Rev. D} {\bf 79} (2009)  086011}, \href{http://arxiv.org/abs/0812.3886}{{\tt arXiv:0812.3886 [hep-th]}}.

\bibitem{Bonifacio:2020xoc}
J.~Bonifacio and K.~Hinterbichler, ``{Bootstrap Bounds on Closed Einstein Manifolds}'', \href{http://dx.doi.org/10.1007/JHEP10(2020)069}{{\em JHEP} {\bf 10} (2020)  069}, \href{http://arxiv.org/abs/2007.10337}{{\tt arXiv:2007.10337 [hep-th]}}.

\bibitem{Bonifacio:2021msa}
J.~Bonifacio, ``{Bootstrap bounds on closed hyperbolic manifolds}'', \href{http://dx.doi.org/10.1007/JHEP02(2022)025}{{\em JHEP} {\bf 02} (2022)  025}, \href{http://arxiv.org/abs/2107.09674}{{\tt arXiv:2107.09674 [hep-th]}}.

\end{thebibliography}\endgroup

\end{document}